\documentclass[preprint,3p,12pt]{elsarticle}
\usepackage{mathrsfs}
\usepackage{amsmath}
\usepackage{stmaryrd}
\usepackage{bbding}
\usepackage{dcolumn}
\usepackage{amsfonts}
\usepackage{amssymb}
\usepackage{psfrag}
\usepackage{wrapfig}
\usepackage{makeidx}
\usepackage{bm}
\usepackage{epsf}
\usepackage{epsfig}
\usepackage{setspace}
\usepackage{epstopdf}
\usepackage{psfrag}
\usepackage{color}
\usepackage{scalerel}  
\usepackage{lineno}
\usepackage{graphicx}
\usepackage{caption}
\usepackage{subcaption}
\usepackage{overpic}
\usepackage{hyperref} 
\usepackage{verbatim}

\begin{document}
\title{An efficient and robust high-order compact ALE gas-kinetic scheme for unstructured meshes}

\author[BNU]{Yibo Wang}
\ead{wangybbnu@163.com}

\author[XJTU]{Xing Ji}
\ead{jixing@xjtu.edu.cn}

\author[BNU]{Liang Pan\corref{cor}}
\ead{panliang@bnu.edu.cn}
\cortext[cor]{Corresponding author}

\address[BNU]{Laboratory of Mathematics and Complex Systems, School of Mathematical Sciences, Beijing Normal University, Beijing, China}
 
\address[XJTU]{State Key Laboratory for Strength and Vibration of Mechanical Structures, Shaanxi Key Laboratory of Environment and Control for Flight Vehicle, School of Aerospace Engineering, Xi'an Jiaotong University, Xi'an, China}

\begin{abstract}
For the arbitrary-Lagrangian-Eulerian (ALE) calculations, the geometric information needs to be calculated at each time step due to the movement of mesh. 
To achieve the high-order spatial accuracy, a large number of matrix inversions are needed, which affect the efficiency of computation dramatically. 
In this paper, an efficient and robust high-order compact ALE gas-kinetic scheme is developed for the compressible moving grids and moving boundary problems. 
The memory-reduction reconstruction \cite{GKS-MR-compact} is used to construct a quadratic polynomial on the target cell, where both structured and unstructured meshes can be used. 
Taking derivatives of the candidate polynomial, the quadratic terms can be obtained by the least square method using the average gradient values of the cell itself and its adjacent cells. 
Moving the quadratic terms to right-hand side of the constrains for cell averaged value, the linear terms of the polynomial can be determined by the least square method as well. 
The gradient compression factor is adopted to suppress the spurious oscillations near discontinuities. 
Combined with the two-stage fourth-order time discretization, a high-order compact gas-kinetic scheme is developed for ALE computation. 
In the process of mesh movement, the inversions of lower order matrix are needed for the least square method, which makes a 7x speedup and improves the efficiency greatly. 
In the computation, the grid velocity can be given by the mesh adaptation method and the cell centered Lagrangian nodal solver.
Numerical examples are presented to evaluate the accuracy, efficiency, robustness and the preservation of geometric conservation law of the current scheme.
\end{abstract}


\maketitle

\section{Introduction}
Considerable progress has been made in the numerical simulations from the Eulerian and Lagrangian methods \cite{Eulerian-Lagrangian}.
Eulerian methods describe the flow motions with time-independent mesh, while the mesh moves with fluid velocity in the Lagrangian methods. 
Integrating the advantages of Eulerian and Lagrangian methods, the arbitrary Lagrangian-Eulerian (ALE) method was originally developed by Hirt \cite{ALE-1}. 
There are two types of ALE methods, i.e. the direct and indirect ALE methods. The indirect ALE methods include three phases \cite{ALE-2}, 
i.e., Lagrangian phase \cite{Lagrangian-c1,Lagrangian-c2,Lagrangian-c3,Lagrangian-s1,Lagrangian-s2,Lagrangian-s3}, rezoning phase \cite{ALE-rezone}, and remapping phase \cite{ALE-remap-1,ALE-remap-2}. 
The solution and mesh are updated in the Lagrangian phase simultaneously. To improve the mesh deformation, 
the mesh is adjusted to the optimal position in the rezoning phase, and the solution is redistributed into the rezoned mesh in the remapping phase.
For the direct ALE methods \cite{ALE3}, the Lagrangian finite volume scheme is based on a space-time coupled conservation formulation.
The grid velocity can be chosen independently from the local fluid velocity, and the remapping phase is not needed.

In the past decades, the gas-kinetic schemes (GKS) based on the Bhatnagar-Gross-Krook (BGK) model \cite{BGK-1,BGK-2} have been
developed systematically for the computations from low speed flows to supersonic ones \cite{GKS-Xu1,GKS-Xu2}. 
The gas-kinetic scheme presents a gas evolution process from the kinetic scale to hydrodynamic scale, 
and  both inviscid and viscous fluxes can be calculated in one framework. Based on the unified coordinate transformation, 
the second-order gas-kinetic scheme was developed under the moving-mesh framework  \cite{GKS-ALE1}.
With the integral form of the fluid dynamic equations, a second-order remapping-free ALE gas-kinetic scheme was developed \cite{GKS-ALE2}, 
which can be considered as a direct ALE method. The grid velocity can be chosen arbitrarily in the flux calculation, 
and the rezoning and remapping stages in the indirect ALE method are avoided.
For high-order accuracy, a one-stage DG-ALE gas-kinetic method was developed, especially for the oscillating airfoil calculations  \cite{GKS-ALE3}. 
However, the one-stage gas evolution model and DG framework become very complicated, 
and the efficiency of the DG-ALE-GKS becomes low due to the severely constrained time step from the DG formulation.
With the two-stage temporal discretization, which was originally developed for the Lax-Wendroff type flow solvers \cite{GRP-high-1,GRP-high-2}, 
a reliable framework was provided to construct gas-kinetic scheme with fourth-order temporal accuracy \cite{GKS-high1}.  
With the non-compact WENO and compact HWENO reconstructions, 
the high-order gas-kinetic schemes are also developed for the ALE computations \cite{GKS-ALE4,GKS-ALE5,GKS-ALE6},
in which the expected order of accuracy is achieved and the geometric conservation law is well preserved. 
For spatial reconstruction \cite{GKS-noncompact,GKS-compact-1,GKS-compact-2}, a quadratic polynomial on the large stencil and several linear polynomials 
on small stencils are combined with non-linear weights to suppress the oscillations in discontinuity region. 
In the computations with steady mesh, the geometric information of the mesh is usually stored in a pre-calculated matrix, 
and the reconstructed variables can be obtained by matrix-vector multiplication in each time step. 
However, in ALE computations, the geometric information needs to be calculated at each time step due to the mesh movement. 
Such calculation includes a large number of matrix inversion operations, which affect the efficiency of computation seriously. 
For the hypersonic flows, the complex flow structures, such as strong shock waves, low-pressure and low-density regions in the flow field, 
are coupled with the process of mesh movement, which brings high demanding for the robustness of numerical schemes. 
In order to overcome these difficulties, it is necessary to develop efficient, robust, and high-order scheme for ALE computations.

Recently, a new memory-reduction compact gas-kinetic scheme is proposed \cite{GKS-MR-compact}, in which a quadratic polynomial can be constructed on the target cell efficiently. 
Taking derivatives of the candidate polynomial, the quadratic terms can be obtained by the least square method using the average gradient values of the cell itself and its adjacent cells. 
Moving the quadratic terms to right-hand side of the constrains for cell averaged value, the linear terms of the polynomial can be determined by the least square method as well. 
As a result, for the two-dimensional case, only one $2\times 2$ coefficient matrix needs to be stored, 
and three times of least square process are needed to construct a quadratic polynomial. 
In this paper, an efficient and robust high-order compact gas-kinetic scheme for ALE computation is developed with the memory-reduction compact reconstruction. 
In the moving mesh procedure, the geometric information needs to be updated for reconstruction at each time step. 
For memory-reduction reconstruction, the $2\times 2$ coefficient matrix is regenerated at each time step inside each cell,  and it is trivial to obtain the inverse matrix. 
Compared with the previous high-order ALE method, the efficiency is improved greatly in the process of mesh movement.
To suppress numerical oscillations in flow with discontinuities, the gradient compression factor is adopted as a limiter at the cell interface. 
To preserve the geometric conservation law, a brief analysis with uniform steady flow is implemented. 
As a result, an extra flux term is added into the gas-kinetic flux, which represents the variation of geometric variable and grid velocity. 
The numerical tests are presented to show the efficiency of current scheme, and 7x speedup can be achieved compared with the previous ALE high-order GKS method \cite{GKS-ALE5}. 
Numerical examples are presented to evaluate the accuracy, efficiency, robustness, and the preservation of geometric conservation law of the current scheme.
In the future, the current scheme will be extended to three-dimensional moving mesh and moving boundary computations.

This paper is organized as follows. In Section 2, the  gas-kinetic scheme in ALE formulation is introduced.
The temporal discretization and geometric conservation law are introduced in Section 3. 
The memory-reduction compact reconstruction is included in Section 4, and the numerical results are presented in section 5. 
The last section is the conclusion.

\section{Gas-kinetic scheme in ALE formulation}
The two-dimensional BGK equation \cite{BGK-1,BGK-2} can be written as
\begin{equation}\label{bgk}
f_t+uf_x+vf_y=\frac{g-f}{\tau},
\end{equation}
where $f$ is the gas distribution function, $g$ is the corresponding equilibrium state and $\tau$ is the collision time.
The collision term satisfies the compatibility condition
\begin{equation}\label{compatibility}
\int \frac{g-f}{\tau}\psi \text{d}\Xi=0,
\end{equation}
where 
$\displaystyle\psi=(1,u,v,\frac{1}{2}(u^2+v^2+\xi^2))^T$, $\text{d}\Xi=\text{d}u\text{d}v\text{d}\xi^1...\text{d}\xi^{K}$, 
$K$ is the number of internal degrees of freedom, i.e., $K=(4-2\gamma)/(\gamma-1)$ for two-dimensional flows, and $\gamma$
is the specific heat ratio.  According to the Chapman-Enskog expansion for BGK equation, the macroscopic governing equations 
can be derived. In the continuum region, the BGK equation can be rearranged and the gas distribution function can be expanded as
\begin{align*}
f=g-\tau D_{\boldsymbol{u}}g+\tau D_{\boldsymbol{u}}(\tau D_{\boldsymbol{u}})g-\tau D_{\boldsymbol{u}}[\tau D_{\boldsymbol{u}}(\tau D_{\boldsymbol{u}})g]+...,
\end{align*}
where $D_{\boldsymbol{u}}=\displaystyle\frac{\partial}{\partial t}+\boldsymbol{u}\cdot \nabla$.  With the zeroth-order truncation
$f=g$, the Euler equations can be obtained. For the first-order truncation
\begin{align*}
f=g-\tau (ug_x+vg_y+g_t),
\end{align*}
the Navier-Stokes equations can be obtained \cite{GKS-Xu1,GKS-Xu2}.

In this paper, the high-order ALE compact gas-kinetic scheme will be constructed in the moving-mesh framework. 
Standing on the moving reference, the BGK equation Eq.\eqref{bgk} becomes
\begin{equation}\label{bgk2}
f_t+(u-U^g)f_x+(v-V^g)f_y =\frac{g-f}{\tau},
\end{equation}
where $\boldsymbol{U}^{g}=(U^{g}, V^{g})$ is the grid velocity. 
Taking moments of Eq.\eqref{bgk2} and integrating with respect to space, the finite volume scheme can be expressed as
\begin{align}\label{finite}
\frac{\text{d}(|\Omega_i|Q_{i})}{\text{d}t}=-\sum_{p \in N(i)}\mathbb{F}_{p}(t),
\end{align}
where $Q_i$ is the cell averaged conservative value over $\Omega_i$, $|\Omega_i|$ is the area of $\Omega_i$.
 $N(i)$ is the set of index for the faces of cell $\Omega_i$, and the time dependent boundary of  $\Omega_i$ is given by
\begin{equation*}
\partial\Omega_i(t)=\bigcup_{p \in N(i)}\Gamma_{p}(t). 
\end{equation*}
The numerical flux is provided by Gaussian quadrature over the cell interface
\begin{align}
\mathbb{F}_p(t)=\left|\Gamma_{p}(t)\right| \sum_{m=1}^2 \omega_{p,m} \boldsymbol{F}_{p,m}(t),
\end{align}
where $|\Gamma_{p}(t)|$ is the length of cell interface and $\omega_{p,m}$ is the quadrature weight.
The numerical flux $\boldsymbol{F}_{p,m}(t)$ can be obtained by taking moments of the gas distribution function
\begin{align}\label{local}
\boldsymbol{F}_{p,m}(t)=\left(
\begin{array}{c}
F^{\rho}_{p,m}   \\
F^{\rho U}_{p,m} \\
F^{\rho V}_{p,m} \\
F^{\rho E}_{p,m} \\
\end{array}\right)=\int\psi f(\boldsymbol{x}_{p,m},t,\boldsymbol{u},\xi) (\boldsymbol{u}-\boldsymbol{U}_{p,m}^g) \cdot \boldsymbol{n}_{p,m}\text{d}\Xi,
\end{align}
where $f(\boldsymbol{x}_{p,m},t,\boldsymbol{u},\xi)$ is the gas distribution function in the global coordinate,
 $\boldsymbol{n}_{p,m}$ is the unit outer normal direction, $\boldsymbol{x}_{p,m}$ is the coordinate of quadrature point and
$\boldsymbol{U}_{p,m}^g$ is the velocity of quadrature point.

To construct the gas distribution function in the local coordinate, the integral solution of the BGK equation is used
\begin{equation}\label{integral1}
f(\boldsymbol{x}_{p,m},t,\boldsymbol{u},\xi)=\frac{1}{\tau}\int_0^t g(\boldsymbol{x}',t',\boldsymbol{u}, \xi)e^{-(t-t')/\tau}\text{d}t'+e^{-t/\tau}f_0(-\boldsymbol{u}t,\xi),
\end{equation}
where $\boldsymbol{u}$ is the particle velocity, $\boldsymbol{x}_{p,m}=\boldsymbol{x}'+\boldsymbol{u}(t-t')$ is the trajectory of particle, 
$f_0$ is the initial gas distribution function, and $g$ is the corresponding equilibrium state. 
According to the integral solution Eq.\eqref{integral1}, 
$f(\boldsymbol{x}_{p,m},t,\boldsymbol{u},\xi)$ at the cell interface can be expressed as
\begin{align}\label{flux}
f(\boldsymbol{x}_{p,m},t,\boldsymbol{u},\xi)=
  & (1-e^{-t/\tau})g_0 + ((t+\tau)e^{-t/\tau}-\tau)(\overline{a}_1u+\overline{a}_2v)g_0\nonumber\\
+ & (t-\tau+\tau e^{-t/\tau}){\bar{A}} g_0\nonumber      \\
+ & e^{-t/\tau}g_r[1-(\tau+t)(a_{1}^{r}u+a_{2}^{r}v)-\tau A^r)]H(u)\nonumber\\
+ & e^{-t/\tau}g_l[1-(\tau+t)(a_{1}^{l}u+a_{2}^{l}v)-\tau A^l)](1-H(u)),
\end{align}
where the coefficients in Eq.\eqref{flux} can be determined by the
reconstructed directional derivatives and compatibility condition
\begin{align*}
\displaystyle 
& \langle a_{1}^{k}\rangle=\frac{\partial Q_{k}}{\partial \boldsymbol{n}_x}, 
  \langle a_{2}^{k}\rangle=\frac{\partial Q_{k}}{\partial \boldsymbol{n}_y}, 
  \langle a_{1}^{k}u+a_{2}^{k}v+A^{k}\rangle=0,\\
\displaystyle 
& \langle\overline{a}_1\rangle=\frac{\partial \overline{Q}}{\partial \boldsymbol{n}_x}, 
  \langle\overline{a}_2\rangle=\frac{\partial \overline{Q}}{\partial \boldsymbol{n}_y}, 
  \langle\overline{a}_1u+\overline{a}_2v+\overline{A}\rangle=0,
\end{align*}
where $k=l,r$, $\boldsymbol{n}_x$, and $\boldsymbol{n}_y$ are local normal and tangential directions, 
and $\langle...\rangle$ is the moment of the equilibrium $g$, which is defined by
\begin{align*}
\langle...\rangle=\int g (...)\psi \text{d}\Xi.
\end{align*}
The spatial derivatives will be obtained by the compact reconstruction,
which will be given in the following section. More details of the
gas-kinetic scheme can be found in \cite{GKS-Xu1}.

In the actual computation, the numerical fluxes are obtained by taking moments of gas distribution function in local coordinate as follows
\begin{align*}
\widetilde{\boldsymbol{F}}_{p,m}(t)=\left(
\begin{array}{c}
F^{\widetilde{\rho}}_{p,m}    \\
F^{\widetilde{\rho U}}_{p,m}  \\
F^{\widetilde{\rho V}}_{p,m}  \\
F^{\widetilde{\rho E}}_{p,m}  \\
\end{array}\right)=
\int\widetilde{u}\left(
\begin{array}{c}
      1        \\
\widetilde{u}  \\
\widetilde{v}  \\
\frac{1}{2}(\widetilde{u}^2+\widetilde{v}^2+\xi^2) \\
\end{array}\right)
f(\boldsymbol{x}_{p,m},t,\widetilde{\boldsymbol{u}},\xi)\text{d}\widetilde{\Xi},
\end{align*}
where the relative particle velocity in the local coordinate is given by
\begin{align*}
\widetilde{\boldsymbol{u}}=(\boldsymbol{u}-\boldsymbol{U}^g_{p,m})\cdot(\boldsymbol{n}_x,\boldsymbol{n}_y)_{p,m}.
\end{align*}
To update the flow variables, the numerical fluxes in global coordinate is needed, and  each component of $\boldsymbol{F}_{p,m}(t)$ can be
given by the combination of $\widetilde{\boldsymbol{F}}_{p,m}(t)$ as follows
\begin{align*}
\left\{\begin{aligned}
F^{\rho}_{p,m} =&F_{p,m}^{\widetilde{\rho}},\\
F^{\rho U}_{p,m}=&U_{p,m}^gF_{p,m}^{\widetilde{\rho}}+a_{11}F_{p,m}^{\widetilde{\rho U}}+a_{12}F_{p,m}^{\widetilde{\rho V}},\\
F^{\rho V}_{p,m}=&V_{p,m}^gF_{p,m}^{\widetilde{\rho}}+a_{21}F_{p,m}^{\widetilde{\rho U}}+a_{22}F_{p,m}^{\widetilde{\rho V}},\\
F^{\rho E}_{p,m}=&F_{p,m}^{\widetilde{\rho E}}+ (a_{11}U_{p,m}^g+a_{21}V_{p,m}^g)F_{p,m}^{\widetilde{\rho U}}
  + (a_{12}U_{p,m}^g+a_{22}V_{p,m}^g)F_{p,m}^{\widetilde{\rho V}}+\frac{1}{2}((U_{p,m}^g)^2+(V_{p,m}^g)^2)F_{p,m}^{\widetilde{\rho}},
\end{aligned} \right.
\end{align*}
where $(a_{ij})$ is denoted as the inverse matrix of $(\boldsymbol{n}_x,\boldsymbol{n}_y)_{p,m}$.

\section{Temporal discretization and geometric conservation law}
Based on the time-dependent flux function of the generalized Riemann problem solver (GRP) 
\cite{GRP-high-1,GRP-high-2} and gas-kinetic scheme \cite{GKS-Xu1,GKS-Xu2}, 
a two-stage fourth-order time-accurate discretization was developed for Lax-Wendroff type flow solvers. 
In this study, the two-stage method is used for gas-kinetic solvers. 
Consider the following time-dependent equation
\begin{align*}
\frac{\partial Q}{\partial t}=\mathcal {L}(Q),
\end{align*}
where $\mathcal {L}$ is an operator for spatial derivative of flux, 
the flow variable $Q^{n+1}$ at $t_{n+1}=t_n+\Delta t$ can be updated with the following formula
\begin{align*}
&Q^*    = Q^n+\frac{1}{2}\Delta t\mathcal{L}(Q^n)+\frac{1}{8}\Delta t^2\frac{\partial}{\partial t}\mathcal{L}(Q^n), \\
 Q^{n+1}=&Q^n+\Delta t\mathcal {L}(Q^n)
        +\frac{1}{6}\Delta t^2\big(\frac{\partial}{\partial t}\mathcal{L}(Q^n)+2\frac{\partial}{\partial t}\mathcal{L}(Q^*)\big).
\end{align*}
It can be proved that for hyperbolic equations the above temporal discretization provides a fourth-order time accurate solution \cite{GRP-high-1,GKS-high1}. 
To implement the two-stage fourth-order method for Eq.\eqref{finite}, 
$\mathcal{L}$ and $\partial_t\mathcal{L}$ at the initial and intermediate stages need to be given. 
For the moving-mesh computation, the temporal discretization above can be extended to Eq.\eqref{finite} by denoting 
\begin{align*}
\frac{\text{d}(|\Omega_i|Q_{i})}{\text{d}t}=\mathcal{L}(\Omega_{i},Q_{i})\triangleq-\sum_{p \in N(i)}\mathbb{F}_{p}(t).
\end{align*}
$\mathcal{L}$ and $\partial_t\mathcal{L}$ at $t^n$ can be given by
\begin{equation}\label{two-stage-2}
\begin{aligned}
\mathcal{L}(\Omega_{i}^n,Q_{i}^n)=&-\sum_{p \in N(i)}\mathbb{F}_{p}^n, \\
\partial_t\mathcal{L}(\Omega_{i}^n,Q_{i}^n)=&-\sum_{p \in N(i)}\partial_t\mathbb{F}_{p}^n.
\end{aligned}
\end{equation}
To determine these coefficients, the time dependent numerical flux $\mathbb{F}_{p}(t)$ in Eq.\eqref{two-stage-2} can be approximate as a linear function
\begin{align}\label{expansion-1}
\mathbb{F}_{p}(t)=\mathbb{F}_{p}^n+ \partial_t\mathbb{F}_{p}^n(t-t_n).
\end{align}
For the gas-kinetic scheme, the gas evolution is a relaxation process from kinetic to hydrodynamic scale through the exponential function, and the corresponding flux is a complicated function of time. 
In order to obtain the time derivatives of the flux function, the flux function should be approximated as a linear function of time within a time interval. 
Integrating Eq.\eqref{expansion-1} over $[t_n, t_n+\Delta t/2]$ and
$[t_n, t_n+\Delta t]$, we have the following two equations
\begin{align}\label{expansion-2}
\begin{cases}
\displaystyle\mathbb{F}_{p}^n\Delta t+\frac{1}{2}\partial_t\mathbb{F}_{p}^n\Delta t^2 =\int_{t_n}^{t+_n\Delta t}\mathbb{F}_{p}(t)  \text{d}t, \\
\displaystyle\frac{1}{2}\mathbb{F}_{p}^n\Delta t+\frac{1}{8}\partial_t\mathbb{F}_{p}^n\Delta t^2 =\int_{t_n}^{t_n+ \Delta t/2}\mathbb{F}_{p}(t)\text{d}t.
\end{cases}
\end{align}
The coefficients $\mathbb{F}_{p}^n$ and $\partial_t\mathbb{F}_{p}^n$
can be obtained by solving the linear system, and
$\mathcal{L}(\Omega_{i}^n,Q_{i}^n)$ and
$\partial_t\mathcal{L}(\Omega_{i}^n,Q_{i}^n)$ can be given by
Eq.\eqref{two-stage-2}.

For the moving-mesh computation, the variation of geometric variable and grid velocity need to be considered
for the numerical flux to preserve the geometric conservation law (GCL). 
Generally, the GCL is referred to as the preservation of uniform flow with arbitrarily 
moving grids for any numerical discretization scheme \cite{GCL-1,GCL-2}. 
In this section, a uniform flow with $\rho = 1$ is adopted for the analysis of GCL. 
Consider a cell interface $\Gamma_p$ at $t=t^n$, which is denoted as $X_1 X_2$. 
$X_1$ and $X_2$ are the vertices on $\Gamma_p$ and the grid velocity of $X_i$ is $\boldsymbol{U}^g_i$, which is constant within a time step. 
The cell interface $\Gamma_p$ can be parameterized as 
\begin{align*}
\boldsymbol{X}(\eta)  &= \sum_{i=1}^{2} \boldsymbol{x}_i \phi_i(\eta), \\
\boldsymbol{U}^g(\eta)&= \sum_{i=1}^{2} \boldsymbol{U}^g_i \phi_i(\eta), 
\end{align*}
where $\eta \in [-1/2,1/2]$, $\boldsymbol{x}_i$ is the location of the vertex $X_i$ and $\phi_i$ is the base function given as follows 
\begin{equation*}
\phi_1 = \frac{1}{2}(1+2\eta), ~\phi_2 = \frac{1}{2}(1-2\eta). 
\end{equation*}
Denote the bilinear functions as follows 
\begin{align*}
  \psi_1 = \frac{1}{4}(1+2\zeta)(1+2\eta), ~\psi_2 = \frac{1}{4}(1+2\zeta)(1-2\eta), \\
  \psi_3 = \frac{1}{4}(1-2\zeta)(1+2\eta), ~\psi_4 = \frac{1}{4}(1-2\zeta)(1-2\eta). 
\end{align*}
The quadrilateral swept by $\Gamma_p$ during a time interval $[t^n,t]$ can be parameterized 
\begin{align*}
\mathbb{X}\left(\zeta,\eta\right) 
&= \sum_{i=1}^{2} \left( \boldsymbol{x}_i\psi_{i}\left( \zeta,\eta  \right) + \left(\boldsymbol{x}_i+\boldsymbol{U}^g_i (t-t^n) \right)\psi_{i+2}\left(\zeta,\eta\right)\right) \\
&= \boldsymbol{X}(\eta) + \frac{1}{2}(1+2\zeta)\boldsymbol{U}^g(\eta)(t-t^n),
\end{align*}
where $(\zeta,\eta) \in [-1/2,1/2]^2$. The swept quadrilateral is denoted as  $X_1 X_2 X_4 X_3$, and the area can be expressed as 
\begin{equation*}
V(t) = \int^{1/2}_{-1/2} \int^{1/2}_{-1/2} | \mathbb{X}_{\zeta}, \mathbb{X}_{\eta} | \text{d}\zeta  \text{d}\eta.
\end{equation*}
With the Gaussian quadrature, the above integration can be expressed as 
\begin{align*}
 V(t) = \sum^2_{m=1} \omega_{p,m} \left( |\boldsymbol{U}^g, \boldsymbol{X}_{\eta}| (t-t^n) + \frac{1}{2}|\boldsymbol{U}^g, \boldsymbol{U}^g_{\eta}| (t-t^n)^2   \right)_{p,m} \Delta \eta. 
\end{align*}

For a steady uniform flow, the density flux $\boldsymbol{F}^\rho_{p,m} = -\boldsymbol{U}^g_{p,m}$, by which the above terms can be transformed as 
\begin{align*}
|\boldsymbol{U}^g, \boldsymbol{X}_{\eta}|_{p,m}
& =  \boldsymbol{U}^g_{p,m} \cdot \boldsymbol{n}_{p,m}\|\boldsymbol{X}_{\eta}\|_p 
  = -\boldsymbol{F}^\rho_{p,m} \|\boldsymbol{X}_{\eta}\|_p,\\
|\boldsymbol{U}^g, \boldsymbol{U}^g_{\eta}|_{p,m}
& =   \boldsymbol{U}^g_{p,m} \cdot \boldsymbol{n}'_{p,m} \|\boldsymbol{U}^g_{\eta}\|_p 
  = -(\boldsymbol{F}')^\rho_{p,m} \|\boldsymbol{U}^g_{\eta}\|_p. 
\end{align*}
where the local normal direction $\boldsymbol{n}_{p,m}$ and $\boldsymbol{n}'_{p,m}$ are determined by
\begin{align*}
  \boldsymbol{n}_{p,m}  &= (\boldsymbol{X}_{\eta}   / ||\boldsymbol{X}_{\eta}||_p)_{\perp}, \\
  \boldsymbol{n}'_{p,m} &= (\boldsymbol{U}^g_{\eta} / ||\boldsymbol{U}^g_{\eta}||_p)_{\perp}.
\end{align*}
Thus, the total numerical flux in Eq.\eqref{expansion-1} can be modified as
\begin{equation*}
  \mathbb{F}_p(t) = \sum^2_{m=1} \omega_{p,m} (\boldsymbol{F}_{p,m} ||\boldsymbol{X}_\eta||_p 
                  + \boldsymbol{F}'_{p,m} ||\boldsymbol{U}^g_{\eta}||_p (t-t^n)), 
\end{equation*}
which is consistent with the result in \cite{GKS-ALE4}. 
The same procedures in Eq.\eqref{expansion-1} and Eq.\eqref{expansion-2} need to be implemented for the two-stage fourth-order method. 
As could be seen in the following accuracy tests, both order of accuracy and GCL are well preserved by current scheme.

\section{Memory-reduction compact reconstruction} 
Different from the traditional Riemann solvers, the gas-kinetic scheme provides a time-dependent gas distribution function by Eq.\eqref{flux}. 
Meanwhile, the macroscopic conservative variables can be obtained by taking moments of the distribution function as well 
\begin{align}\label{point}
Q(\boldsymbol{x}_{p,m},t)=\int\boldsymbol\psi f(\boldsymbol{x}_{p,m},t,\boldsymbol{u},\xi)\text{d}\Xi, 
\end{align}
where $f(\boldsymbol{x}_{p,m},t,\boldsymbol{u},\xi)$ is the gas distribution function at $m$-th Gaussian point on the  interface $p$. 
According to the Gauss's theorem, the cell-averaged gradient of the flow variable $Q$ can be calculated as follows 
\begin{equation*}
\begin{split}
|\Omega_i|(\nabla Q)_i(t)&=\iint_{\Omega_i}\nabla Q(t)\text{d}V=\int_{\partial\Omega_i}Q(t)\boldsymbol{\tau}\text{d}\Gamma \\
&=\sum_{p\in N(i)} \Big( \sum_{m=1}^2 \omega_{p,m} \Big(\int\boldsymbol\psi f(\boldsymbol{x}_{p,m},t,\boldsymbol{u},\xi)\text{d}\Xi\Big) \boldsymbol{\tau} |\Gamma_{p}| \Big),
\end{split}
\end{equation*}
where $\Omega_i$ is arbitrary cell in computational mesh, $(\nabla Q)_i$ is the cell averaged gradient of the flow variable over 
cell $\Omega_i$, $\nabla Q$ is the distribution of flow gradient, $\boldsymbol{\tau}$ takes $\boldsymbol{n_x}$ and $\boldsymbol{n_y}$, which are the unit directions of local coordinate on the cell interface. 
The macroscopic variables $Q_{p,m}^{n+1}$ at a cell interface can be obtained by taking moments of $f_{p,m}^{n+1}$ and the cell interface values can be used for the reconstruction at the beginning of next time step. 
In order to utilize the two-stage fourth-order temporal discretization for the gas distribution function, the second-order gas-kinetic solver is needed.  
In the previous study, the gas distribution function is recommended to be updated by 
\begin{equation}\label{two-stage-3}
\begin{split}
&f^*=f^n+\frac{1}{2}\Delta tf_t^n, \\
&f^{n+1}=f^n+\Delta t f_{t}^*.
\end{split}
\end{equation}
where the subscripts $m,p$ are omitted. 
It is relatively simple to be implemented, but the numerical tests show that only third-order accuracy can be achieved.  
With Gaussian quadrature rule, the cell-averaged gradient $(\nabla Q)_i$ at $t=t^n$ can be obtained. 
It means that the computation of cell-averaged gradient $(\nabla Q)_i$ can be turned to calculate the point value $Q_{p,m}$ on the cell interface in the local orthogonal coordinate.

In the traditional HWENO method, the quadratic polynomial and linear polynomials need to be constructed at every time step. 
For the stationary mesh, the matrix for geometric information is stored at the initial step. 
However, for the ALE computation, with the moving mesh, 
the matrix needs to be calculated at every time step, which is very time consuming. 
To improve the efficiency, the memory-reduction compact reconstruction \cite{GKS-MR-compact} is introduced
for both structured and unstructured meshes. 
For the target cell $\Omega_{i}$, the big stencil for the compact reconstruction is selected as 
\begin{align*}
  S_i^{HWENO}&=\{Q_{i},Q_{i_1},\dots, Q_{i_p},\nabla Q_{i},\nabla Q_{i_1},\dots,\nabla Q_{i_p}\},
\end{align*}
where $Q_{i_p}$ and $\nabla Q_{i_p}$ are the cell averaged conservative variable and cell averaged gradient of the neighboring cell $\Omega_{i_p}$.
For simplicity, the big stencil is rearranged as 
\begin{align*}
  S_i^{HWENO}&=\{Q_0,Q_1,…,Q_{p},\nabla Q_0,...,\nabla Q_{p}\}. 
\end{align*}
Based on the big stencil, a quadratic polynomial can be constructed as
\begin{align*}
  P_0(\boldsymbol{x})&=Q_{0}+\sum_{|\boldsymbol n|=1}^2a_{\boldsymbol n}p_{\boldsymbol n}(\boldsymbol{x}), 
\end{align*}
where $Q_{0}$ is the cell averaged conservative variables over $\Omega_{0}$, 
the multi-index $\boldsymbol n=(n_1, n_2)$ and $|\boldsymbol n|=n_1+n_2$. 
The base function $p_{\boldsymbol n}(\boldsymbol{x})$ is defined as
\begin{align*}
\displaystyle p_{\boldsymbol n}(\boldsymbol{x}) = 
(x-x_0)^{n_1} (y-y_0)^{n_2}-\frac{1}{\left|\Omega_{0}\right|}\iint_{\Omega_{0}}(x-x_0)^{n_1}(y-y_0)^{n_2}\text{d}V.
\end{align*}
To determine the coefficients of $P_0(\boldsymbol{x})$,  the derivatives are given by
\begin{align*}
\partial_x P_0(\boldsymbol{x})&=a_1 +a_{11}(x-x_0)+a_{12}(y-y_0),\\
\partial_y P_0(\boldsymbol{x})&=a_2 +a_{12}(x-x_0)+a_{22}(y-y_0),
\end{align*}
where $\partial_x P_0(\boldsymbol{x}),\partial_y P_0(\boldsymbol{x})$ represent the derivatives in $x$, $y$ directions of $P_0(\boldsymbol{x})$. 
Meanwhile, the distribution of cell-averaged slopes can be written as linear polynomials as follows 
\begin{align*}
L^1_x(\boldsymbol{x})&=b_0 +b_1(x-x_0)+b_2(y-y_0),\\
L^1_y(\boldsymbol{x})&=c_0 +c_1(x-x_0)+c_2(y-y_0),
\end{align*}
where the following constraints are satisfied
\begin{equation*}
\begin{split}
\frac{1}{\left|\Omega_{k}\right|}\iint_{\Omega_{k}}L^1_x(\boldsymbol{x})\text{d}V
&=(Q_x)_{k},~(Q_{x})_{k}\in S_i^{HWENO}, \\
\frac{1}{\left|\Omega_{k}\right|}\iint_{\Omega_{k}}L^1_y(\boldsymbol{x})\text{d}V
&=(Q_y)_{k},~(Q_{y})_{k}\in S_i^{HWENO}.
\end{split}
\end{equation*}
With the least-square process, the coefficients in $L^1_x(\boldsymbol{x})$ and $L^1_y(\boldsymbol{x})$ can be determined.
Compared with $\partial_x P_0(\boldsymbol{x})$ and $\partial_y P_0(\boldsymbol{x})$, the quadratic terms in $P_0(\boldsymbol{x})$ can be obtained as 
\begin{align*}
& a_{11} =  b_1,  a_{22} =  c_2,  a_{12} = (b_2 +c_1)/2. 
\end{align*}
To fully determine the quadratic polynomial $P_0(\boldsymbol{x})$, the constraints for the cell averaged variable are considered
\begin{equation*} 
\begin{split}
\frac{1}{\left|\Omega_{k}\right|}\iint_{\Omega_{k}}P_0(\boldsymbol{x})\text{d}V&=Q_{k}, ~Q_{k}\in S_i^{HWENO}. \\
\end{split}
\end{equation*}
Moving the quadratic terms to the right hand side, a new linear system of the linear terms can be obtained
\begin{equation*} 
\begin{split}
\iint_{\Omega_{k}}a_{1}(x-x_0)+a_{2}(y-y_0)\text{d}V& = 
\left|\Omega_{k}\right|\Delta Q_{k} - \iint_{\Omega_{k}}(\sum_{|\boldsymbol n|=2}a_{\boldsymbol n}p_{\boldsymbol n}(\boldsymbol{x}))\text{d}V, 
\end{split}
\end{equation*}
where $\Delta Q_{k}= (Q_{k}-Q_{0})$.  Solving the linear system with the least square method, all coefficients of $P_0(x)$ for the big stencil are determined.

With the reconstructed $P_0(\boldsymbol{x})$ inside each cell, the conservative variables and their gradients at the Gaussian quadrature point can be reconstructed at cell interface, 
which are denoted as $Q^{l,r}_{p,m}, (Q^{l,r}_x)_{p,m}, (Q^{l,r}_y)_{p,m}$. Subscripts $l,r$ represent the left and right side of the interface, 
and $p,m$ denote $m$-th Gaussian quadrature point at interface $p$ of the target cell. 
However , the numerical oscillations usually occur near discontinuities, when the linear reconstructed variables and gradients are directly used for flux calculation. 
To deal with discontinuity, the gradient compression factor is adopted. 
The compression factor  \cite{GKS-compact-gcf,GKS-MR-compact} for target cell $\Omega_i$ is given as follows
\begin{equation*}
\alpha_i=\prod_{p\in N(i)} \prod_{m=1}^{2} \alpha_{p,m}.
\end{equation*}
The compression factor $\alpha_{p,m}$ at Gaussian quadrature point can be calculated by 
\begin{align*}
\alpha_{p,m}=\frac{1}{1+A^2}, 
\end{align*}
where
\begin{align*}
A=\frac{\left|p^l-p^r\right|}{p^l}+\frac{\left|p^l-p^r\right|}{p^r}+\left(\mathrm{Ma}_n^l-\mathrm{Ma}_n^r\right)^2+\left(\mathrm{Ma}_t^l-\mathrm{Ma}_t^r\right)^2, 
\end{align*}
and $p$ is pressure, $Ma_n$ and $Ma_t$ are the Mach numbers defined by normal and tangential velocity. 
With the gradient compression factor, the updated cell-averaged slopes can be modified as
\begin{equation*}
  \widetilde{\nabla Q}^{n+1}_{i} = \alpha_i \nabla Q^{n+1}_{i}.
\end{equation*}
The compression factor can be adopted to directly limit the reconstructed values and gradients on cell interface as a limiter. 
Denote $Q_l$ the cell-averaged variables in the left cell of target face. 
With $\alpha_{p,m}$ at Gaussian quadrature point, the reconstructed values on the left side can be modified as
\begin{align*}
\widetilde{Q^{l}}_{p,m}=Q_l + \alpha_{p,m} (Q^{l}_{p,m}-Q_l), 
\end{align*}
and  
\begin{align*}
(\widetilde{Q^{l}_x})_{p,m}=\alpha_{p,m} (Q^{l}_x)_{p,m}, \\
(\widetilde{Q^{l}_y})_{p,m}=\alpha_{p,m} (Q^{l}_y)_{p,m}. 
\end{align*}
The right side values can be modified similarly. For the numerical examples,  the compression factor works well for the cases with strong shocks, which verifies the robustness of the limiting process. 
However, due to the simple limiting strategy, the compression factor seems to be dissipative in the smooth regions. 
In the future, the robust and efficient limiters still need to be investigated for the compact gas-kinetic scheme.

\section{Numerical tests}
In this section, the numerical tests for smooth flows and flows with strong discontinuity will be presented to validate our numerical scheme. 
For the inviscid flow, the collision time $\tau$ takes 
\begin{align*}
\tau=\epsilon \Delta t+C\displaystyle|\frac{p^l-p^r}{p^l+p^r}|\Delta t,
\end{align*}
where $\varepsilon=0.1$ and $C=1$, $p^l$ and $p^r$ denote the pressure on the left and right sides of the cell interface. 
Without spatial statement, the ratio of specific heats takes $\gamma=1.4$.

\begin{table}[!h]{}
\begin{center}
\def\temptablewidth{0.75\textwidth}
{\rule{\temptablewidth}{1.0pt}}
\begin{tabular*}{\temptablewidth}{@{\extracolsep{\fill}}c|cc|cc}
mesh     & $L^1$ error &  order   & $L^2$ error &  order  \\
 \hline
 $16^2$   &  2.0365E-02 &  ~       &  1.1254E-02 &  ~      \\
 $32^2$   &  2.6392E-03 &  2.9480  &  1.4622E-03 &  2.9442 \\
 $64^2$   &  3.3206E-04 &  2.9906  &  1.8404E-04 &  2.9901 \\
 $128^2$  &  4.1568E-05 &  2.9979  &  2.3029E-05 &  2.9985 \\
 $256^2$  &  5.1968E-06 &  2.9998  &  2.8787E-06 &  3.0000 \\
\end{tabular*}
 {\rule{\temptablewidth}{1.0pt}}
 \end{center}
 \caption{\label{accuracy-1} Advection of density perturbation: $L^1$ and $L^2$ errors and orders of accuracy with the stationary mesh at $t=2$.}
\end{table}

\subsection{Accuracy tests}
\begin{figure}[!htb]
\centering
\includegraphics[width=0.475\textwidth]{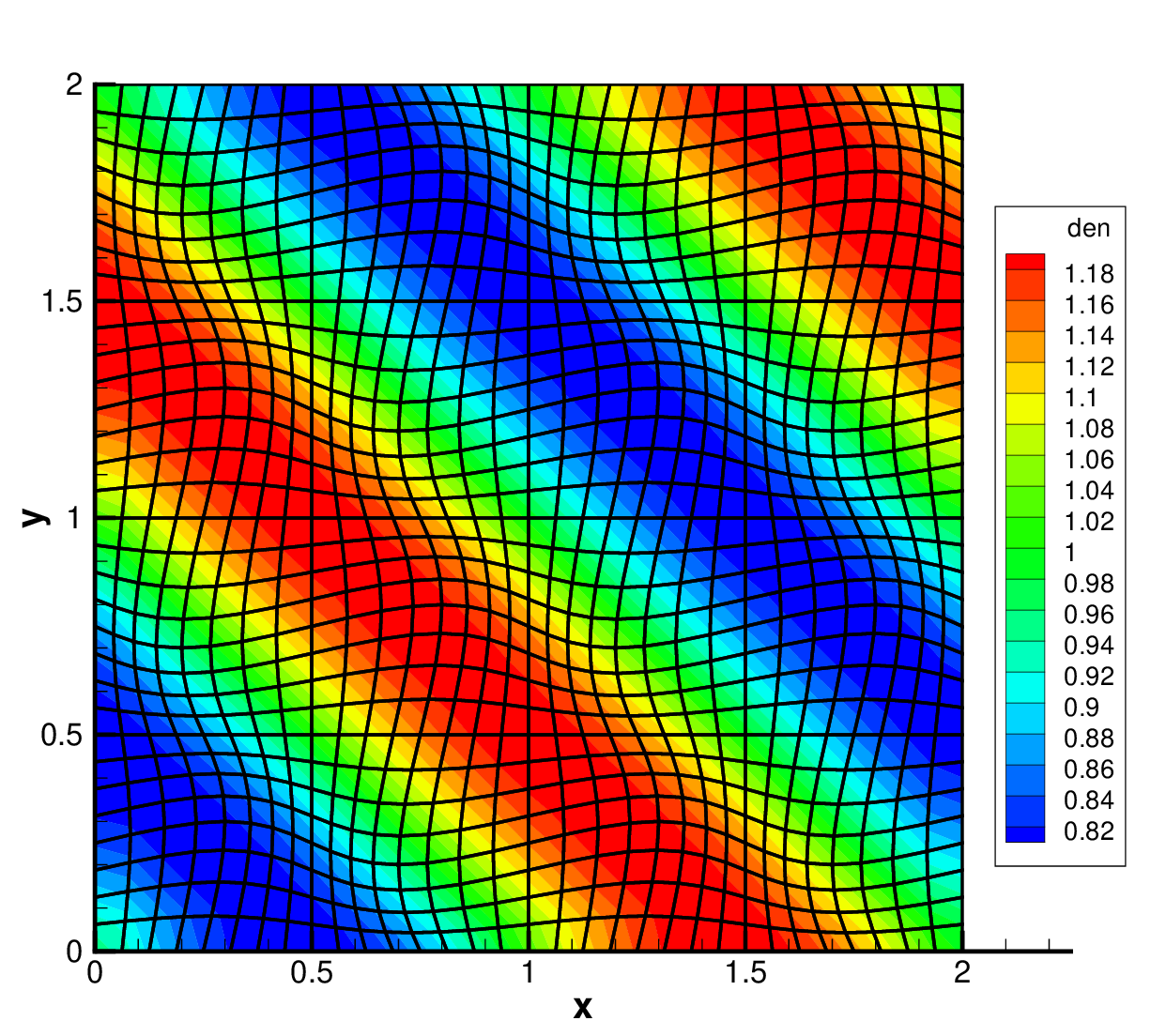} 
\includegraphics[width=0.475\textwidth]{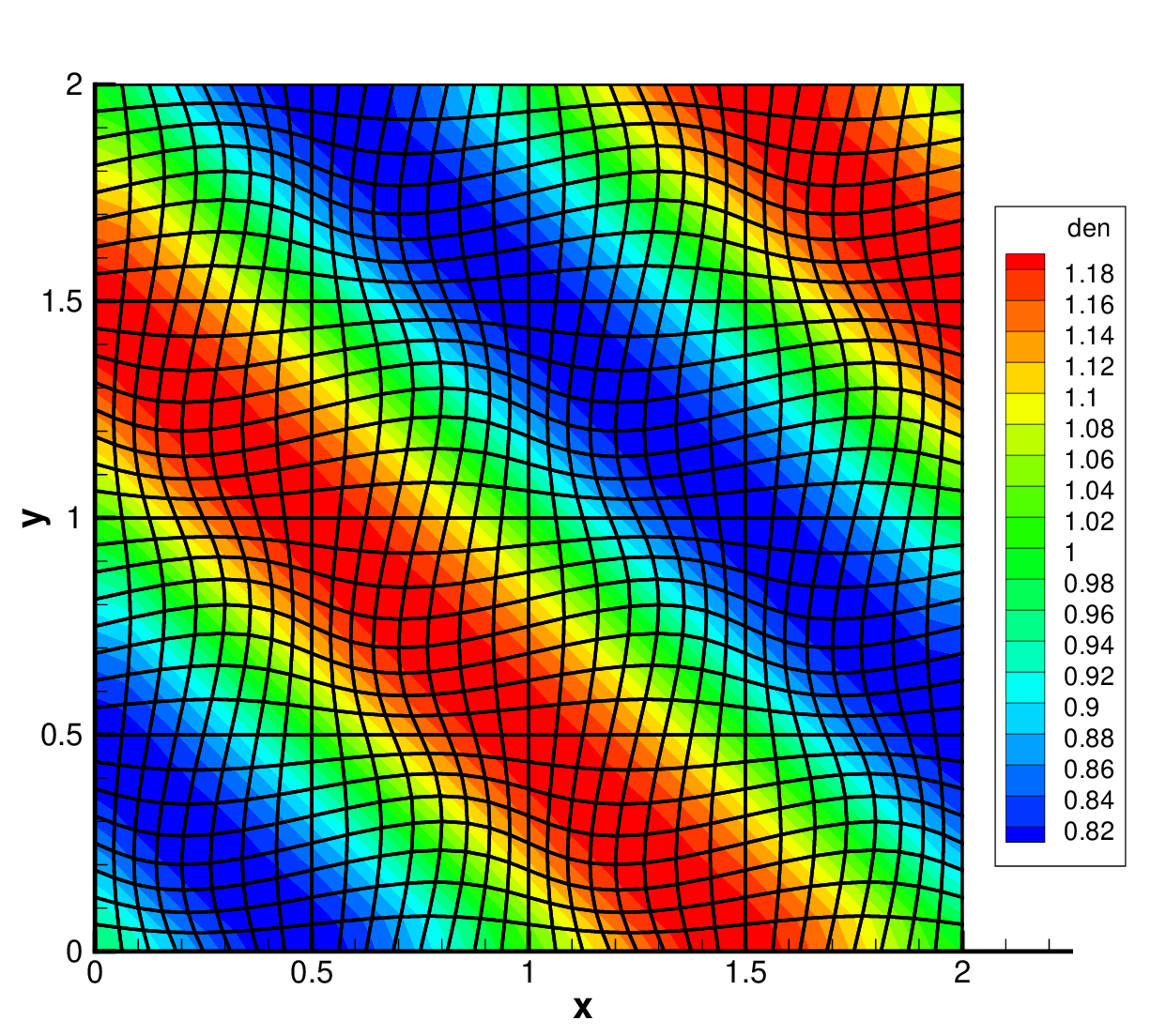}
\caption{\label{Accuracy-mesh}  Advection of density perturbation: the mesh and density distribution for the moving mesh Type-2 at $t=0.5$ (left) and $t=0.5$ (right).}
\end{figure}

The first case for accuracy test is the advection of density perturbation. 
The computational domain is $[0,2]\times[0,2]$, and the initial condition is set as 
\begin{align*}
\rho(x,y)=1+0.2\sin(\pi (x+y)),\ \  U(x,y)=1,\ \  V(x,y)=1,\ \ \  p(x,y)=1.
\end{align*}
The periodic boundary conditions are adopted at all boundaries, and the analytic solution is given by
\begin{align*}
\rho(x,y,t)=1+0.2\sin(\pi(x+y-2t)),\ \  U(x,y,t)=1,\ \  V(x,y,t)=1,\ \  p(x,y,t)=1.
\end{align*}
To test the accuracy for current reconstruction, the stationary uniform mesh with mesh size $h = 2/N$ is used. 
The $L^1$ and $L^2$ errors and orders at $t=2$ are presented in Tab.\ref{accuracy-1}, and 
the expected order of accuracy can be achieved on stationary mesh. 
To validate the order of accuracy with moving-mesh, the following three time-dependent moving meshes are considered
\begin{align*}
\text {Type-1:}& 
\begin{cases}
x(t)=x_0+0.05 \sin \pi t \sin \pi x_0 \sin \pi y_0, \\
y(t)=y_0+0.05 \sin \pi t \sin \pi x_0 \sin \pi y_0,
\end{cases}\\
\text {Type-2:}& 
\begin{cases}
x(t)=x_0+0.05 \sin \pi t \sin 2 \pi x_0 \sin 2 \pi y_0, \\
y(t)=y_0+0.05 \sin \pi t \sin 2 \pi x_0 \sin 2 \pi y_0,
\end{cases} \\
\text {Type-3:}& 
\begin{cases}
x(t)=x_0+0.05 \sin \pi t \sin \pi x_0, \\
y(t)=y_0+0.05 \sin \pi t \sin \pi y_0,
\end{cases}    
\end{align*}
where $(x_0,y_0)$ denote the initial uniform mesh with mesh size $h = 2/N$. 
The grid velocity can obtained by taking temporal derivatives of $x(t)$ and $y(t)$. 
The mesh and density distribution at $t=0.5$ and $t=1.5$ for Type-2 moving mesh is shown in Fig.\ref{Accuracy-mesh}. 
The $L^1$ and $L^2$ errors and order of accuracy at $t = 2$ are presented in Tab.\ref{accuracy-2}, 
and expected order of accuracy can be obtained with the moving meshes as well.

\begin{table}[!h]
\begin{center}
\def\temptablewidth{0.8\textwidth}
{\rule{\temptablewidth}{1.0pt}}
\begin{tabular*}{\temptablewidth}{@{\extracolsep{\fill}}c|c|cc|cc}
~               &  mesh    & $L^1$ error &  order   & $L^2$ error &  order  \\
\hline 
$\text{Type-1}$	& $16^2$   &  2.1682E-02 &  ~       &  1.1991E-02 &  ~  \\
~               & $32^2$   &  2.8331E-03 &  2.9361  &  1.5723E-03 &  2.9310 \\
~               & $64^2$   &  3.5750E-04 &  2.9864  &  1.9834E-04 &  2.9868 \\
~               & $128^2$  &  4.4762E-05 &  2.9976  &  2.4830E-05 &  2.9978 \\
~               & $256^2$  &  5.5962E-06 &  2.9998  &  3.1040E-06 &  2.9999 \\
\hline 
\hline 
$\text{Type-2}$	& $16^2$   &  2.3245E-02 &  ~       &  1.2974E-02 &  ~  \\
~               & $32^2$   &  3.3004E-03 &  2.8162  &  1.8410E-03 &  2.8170 \\
~               & $64^2$   &  4.2679E-04 &  2.9510  &  2.3918E-04 &  2.9443 \\
~               & $128^2$  &  5.3783E-05 &  2.9883  &  3.0184E-05 &  2.9862 \\
~               & $256^2$  &  6.7334E-06 &  2.9978  &  3.7803E-06 &  2.9972 \\
\hline 
\hline 
$\text{Type-3}$	& $16^2$   &  2.1079E-02 &  ~       &  1.1807E-02 &  ~  \\
~               & $32^2$   &  2.7750E-03 &  2.9252  &  1.5564E-03 &  2.9233 \\
~               & $64^2$   &  3.5109E-04 &  2.9826  &  1.9725E-04 &  2.9801 \\
~               & $128^2$  &  4.4959E-05 &  2.9652  &  2.5555E-05 &  2.9484 \\
~               & $256^2$  &  6.3330E-06 &  2.8276  &  3.7852E-06 &  2.7551 \\
\end{tabular*}
{\rule{\temptablewidth}{1.0pt}}
\end{center}
\caption{\label{accuracy-2}  Advection of density perturbation:  $L^1$ and $L^2$ errors and orders of accuracy with the 
moving mesh Type-1, Type-2 and Type-3 at $t=2$. }
\end{table}

The geometric conservation law (GCL) is also tested. Similar to the advection of density perturbation, 
the computational domain is $[0,2]\times[0,2]$, and the periodic boundary conditions are adopted at all boundaries. 
The uniform initial condition is set as 
\begin{align*}
(\rho, U, V, p) = (1, 1, 1, 1).
\end{align*}
Three time-dependent meshes, i.e., Type-1, Type-2 and Type-3 are considered. 
The $L^1$ and $L^2$ errors at $t=2$ with different moving meshes are presented in Tab.\ref{accuracy-gcl}. 
For the mesh velocity given by Type-1 and Type-2, the integrated extra flux is always zero, 
and the geometric conservation law is satisfied automatically. 
For Type-3 moving mesh, the geometric variation has to be included in the numerical fluxes. 
As shown in Tab.\ref{accuracy-gcl}, both $L^1$ and $L^2$ errors are reduced to machine zero, which implies that current ALE 
method well preserves the geometric conservation law. 

\begin{table}[!h]
\begin{center}
\def\temptablewidth{0.5\textwidth}
{\rule{\temptablewidth}{1.0pt}}
\begin{tabular*}{\temptablewidth}{@{\extracolsep{\fill}}c|c|c c}          
                & mesh    & $L^1$ error & $L^2$ error\\
\hline
$\text{Type-1}$ &$8^2$    & 5.7468E-14  & 4.0275E-14 \\
                & $16^2$  & 2.6512E-14  & 1.9265E-14 \\
                & $32^2$  & 1.2974E-14  & 8.8872E-15 \\
                & $64^2$  & 1.4285E-14  & 9.3422E-15 \\
                & $128^2$ & 2.1494E-14  & 1.4008E-14 \\
\hline
\hline 
$\text{Type-2}$ &$8^2$    & 2.6522E-13  & 1.6063E-13 \\
                & $16^2$  & 1.1339E-13  & 7.4030E-14 \\
                & $32^2$  & 4.9109E-14  & 3.4439E-14 \\
                & $64^2$  & 3.0853E-14  & 2.1049E-14 \\
                & $128^2$ & 2.8271E-14  & 1.8343E-14 \\
\hline
\hline 
$\text{Type-3}$ &$8^2$    & 6.0299E-15  & 3.6800E-15 \\
                & $16^2$  & 8.7153E-15  & 5.3347E-15 \\
                & $32^2$  & 1.2369E-14  & 7.7087E-15 \\
                & $64^2$  & 1.8025E-14  & 1.1366E-14 \\
                & $128^2$ & 2.8191E-14  & 1.8078E-14 \\
\end{tabular*}
{\rule{\temptablewidth}{1.0pt}}
\end{center}
\caption{\label{accuracy-gcl} Geometric conservation law:  $L^1$ and $L^2$ errors with moving mesh Type-1, Type-2 and Type-3.  }
\end{table}

The second one is the isotropic vortex propagation, which is a non-linear case for accuracy test. 
The computational domain is $[0,10]\times[0,10]$, and the periodic boundary conditions are adopted at all boundaries. 
The mean flow is $(\rho, U, V, p) = (1, 1, 1, 1)$, and an isotropic vortex is added by perturbation in $U, V$ and temperature $T$, which is given by
\begin{equation*}
\begin{gathered}
(\delta U, \delta V)=\frac{\epsilon}{2 \pi} e^{\frac{(1-r^2)}{2}}(-y, x), \\
\delta T=-\frac{(\gamma-1) \epsilon^2}{8 \gamma \pi^2} e^{1-r^2}, \delta S=0, 
\end{gathered}
\end{equation*}
where $r^2 = x^2 + y^2$, and the vortex strength $\epsilon = 5$. 
The exact solution is that the given perturbation propagates with the velocity $(U, V) = (1, 1)$. 
To test the accuracy with moving mesh, the following time-dependent mesh is considered
\begin{align*}
\text{Type-4:}& 
\begin{cases}
 x(t)=x_0+0.05 \sin 0.2 \pi t \sin \pi x_0 \sin \pi y_0, \\
 y(t)=y_0+0.05 \sin 0.2 \pi t \sin \pi x_0 \sin \pi y_0, 
\end{cases}    
\end{align*}
where $(x_0,y_0)$ denote the initial uniform mesh with mesh size $h = 10/N$. 
The $L^1$ and $L^2$ error and order of accuracy at $t = 10$ for both stationary 
and moving meshes are presented in Tab.\ref{accuracy-3}. 
The expected accuracy is obtained for both stationary and moving meshes. 

\begin{table}[!h]
\begin{center}
\def\temptablewidth{0.85\textwidth}
{\rule{\temptablewidth}{1.0pt}}
\begin{tabular*}{\temptablewidth}{@{\extracolsep{\fill}}c|c|cccc}
 ~              & mesh    & $L^1$ error &  order & $L^2$ error &  order \\
\hline 
stationary mesh & $32^2$  &  3.2080E-01 &  ~     &  8.1228E-02 & ~      \\
  	            & $64^2$  &  5.6194E-02 & 2.5132 &  1.4194E-02 & 2.5166 \\
	              & $128^2$ &  7.9427E-03 & 2.8227 &  1.9909E-03 & 2.8338 \\
	              & $256^2$ &  1.0870E-03 & 2.8693 &  2.5948E-04 & 2.9398 \\
\hline 
\hline 
moving mesh     & $32^2$  &  3.2239E-01 & ~      &  8.1527E-02 & ~      \\
  	            & $64^2$  &  5.7629E-02 & 2.4839 &  1.4449E-02 & 2.4963 \\
  	            & $128^2$ &  8.2881E-03 & 2.7977 &  2.0422E-03 & 2.8227 \\
  	            & $256^2$ &  1.1373E-03 & 2.8654 &  2.6649E-04 & 2.9379 \\
\end{tabular*}
{\rule{\temptablewidth}{1.0pt}}
\end{center}
\caption{\label{accuracy-3} Isotropic vortex propagation: $L^1$ and $L^2$ errors and orders of accuracy with stationary mesh and moving-mesh at $t=10$. }
\end{table}

\subsection{Efficiency comparison}

The efficiency comparison of the current scheme and the previous ALE high-order GKS method \cite{GKS-ALE4} is presented. 
With the memory-reduction reconstruction process, only a $2\times2$ coefficient matrix is regenerated at each time step inside each cell, and it is trivial to get the inverse matrix. 
The large number of matrix inversion operations in the traditional high-order ALE methods can be omitted, which makes the current scheme very simple and efficient. 
The advection of density perturbation with Type-$1$ moving mesh is tested for execution time within one period. 
Both codes are compiled and run with the Intel Core i7-9700 CPU, and the execution time with different meshes are given in terms of seconds. 
As shown in Tab.\ref{efficiency},  7x speedup can be achieved, which validates the efficiency of the current high-order compact ALE GKS.

\begin{table}[!h]
\begin{center}
\def\temptablewidth{0.8\textwidth}
{\rule{\temptablewidth}{1.0pt}}
\begin{tabular*}{\temptablewidth}{@{\extracolsep{\fill}}c|c|c|c|c}        
mesh    & time step   & current scheme & previous scheme & speedup \\
\hline
$32^2$  & $5.00\times 10^{-3}$ & 2   & 14   & 7.0 \\
$64^2$  & $2.50\times 10^{-3}$ & 15  & 109  & 7.3 \\
$128^2$ & $1.25\times 10^{-3}$ & 109 & 1067 & 7.8 \\
\end{tabular*}
{\rule{\temptablewidth}{1.0pt}}
\end{center}
\caption{\label{efficiency} Efficiency comparison: the total execution time (in seconds) and speedup for current and previous scheme for advection of density perturbation at $t=2$. }
\end{table}

\begin{figure}[!htb]
\centering
\includegraphics[width=0.475\textwidth]{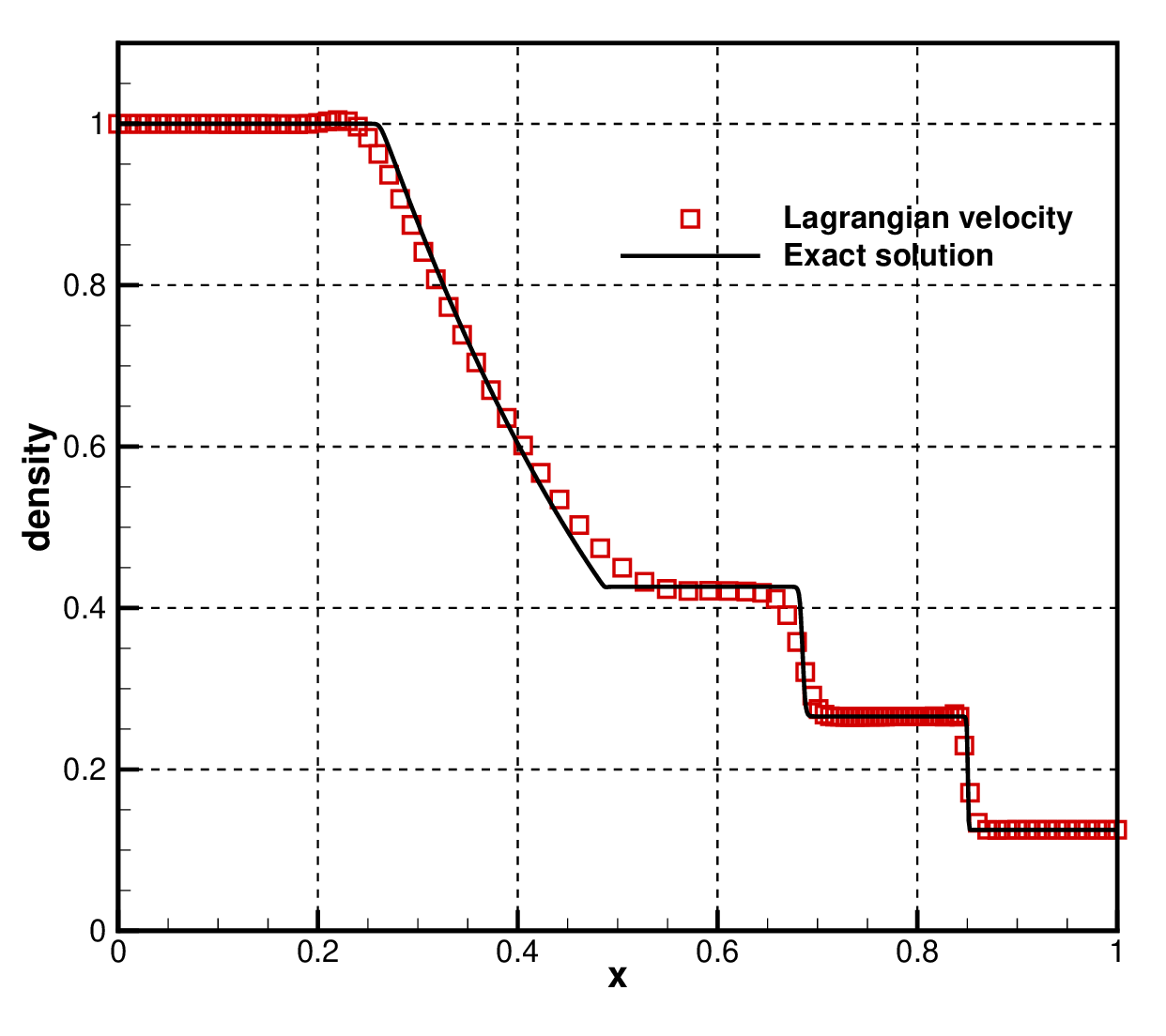} 
\includegraphics[width=0.475\textwidth]{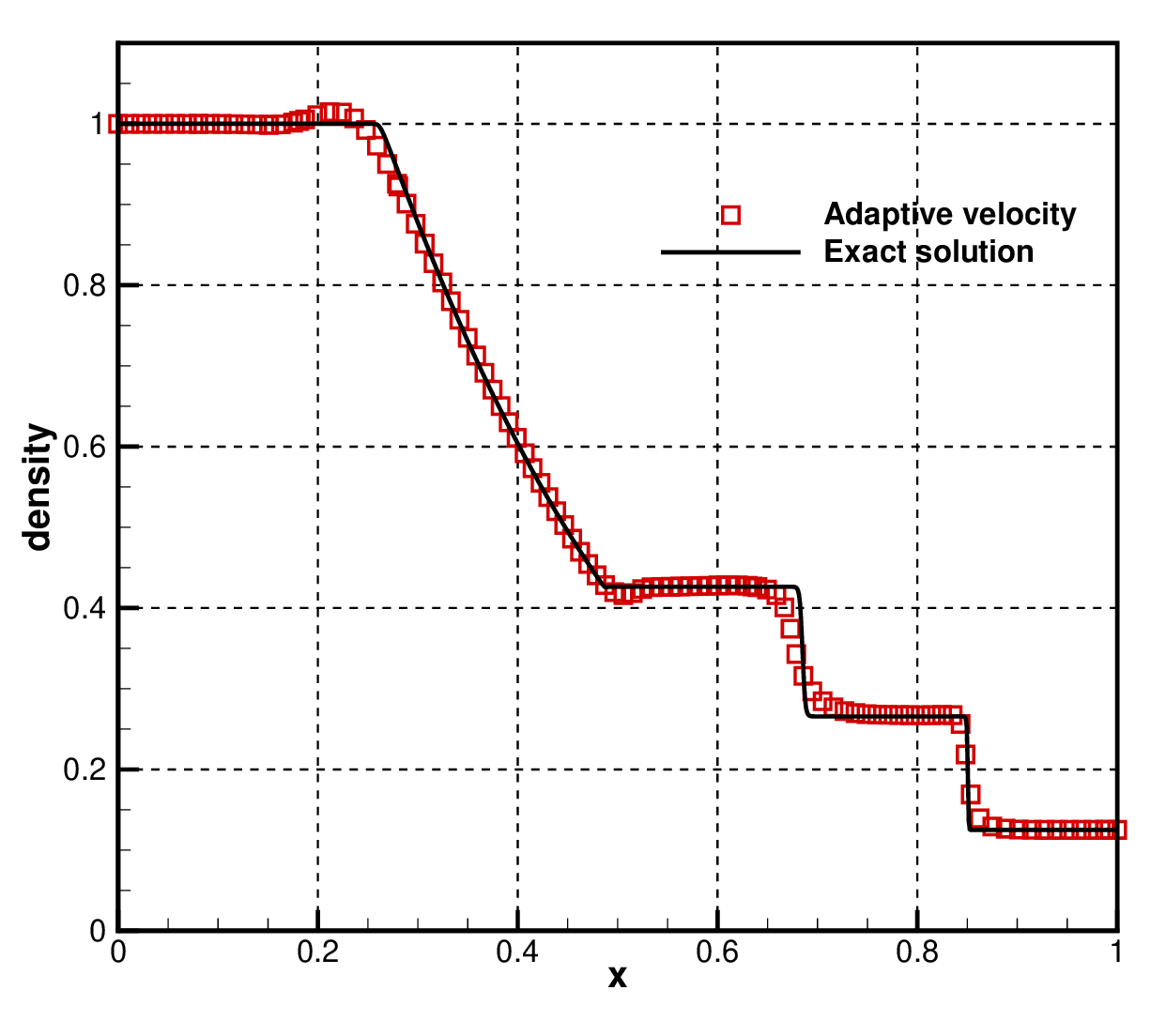}
\caption{\label{sod-density} One-dimensional Riemann  problem: the density distribution with Lagrangian velocity (left) and adaptation velocity (right) at $t=0.2$ for Sod problem.}
\centering
\includegraphics[width=0.475\textwidth]{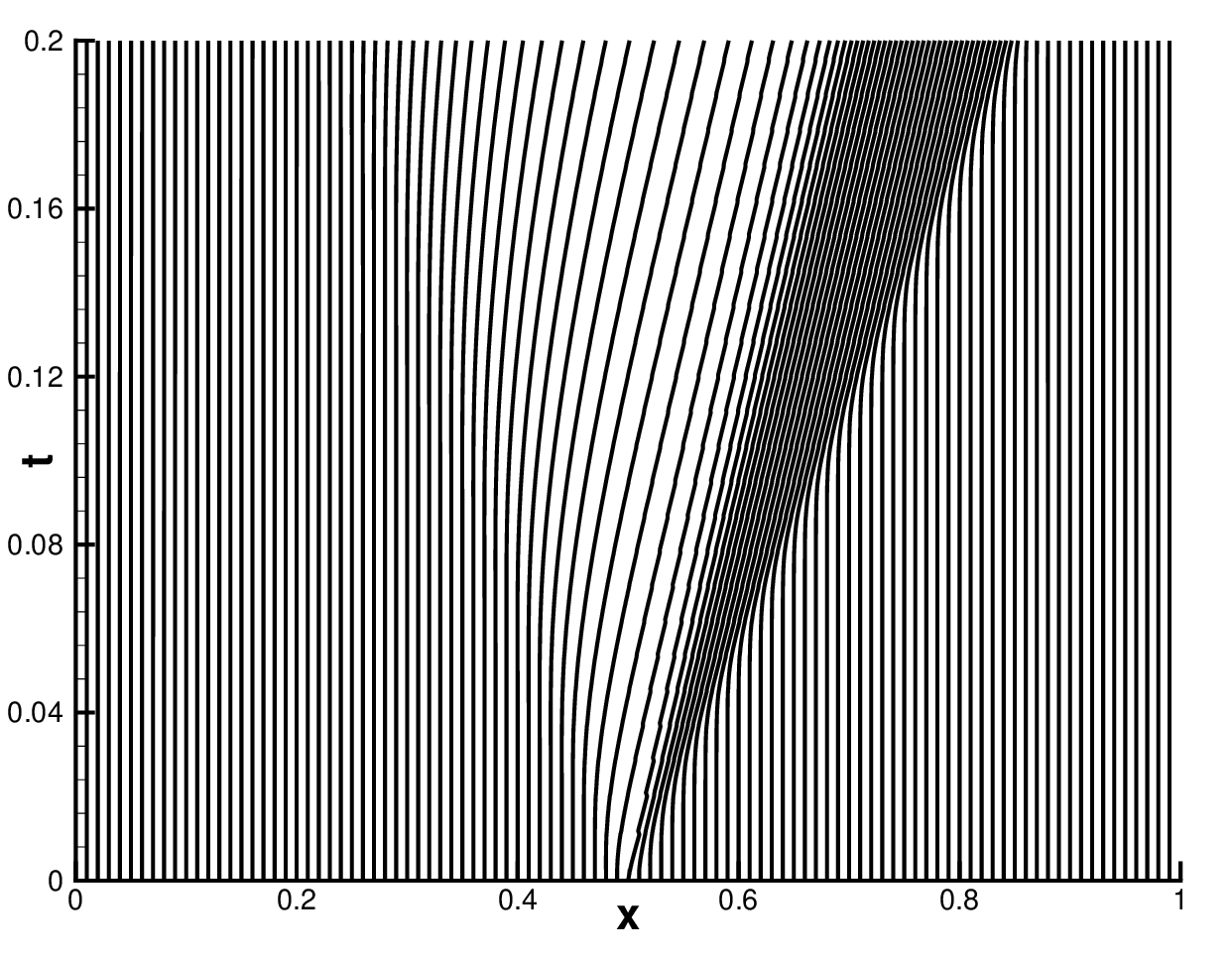}
\includegraphics[width=0.475\textwidth]{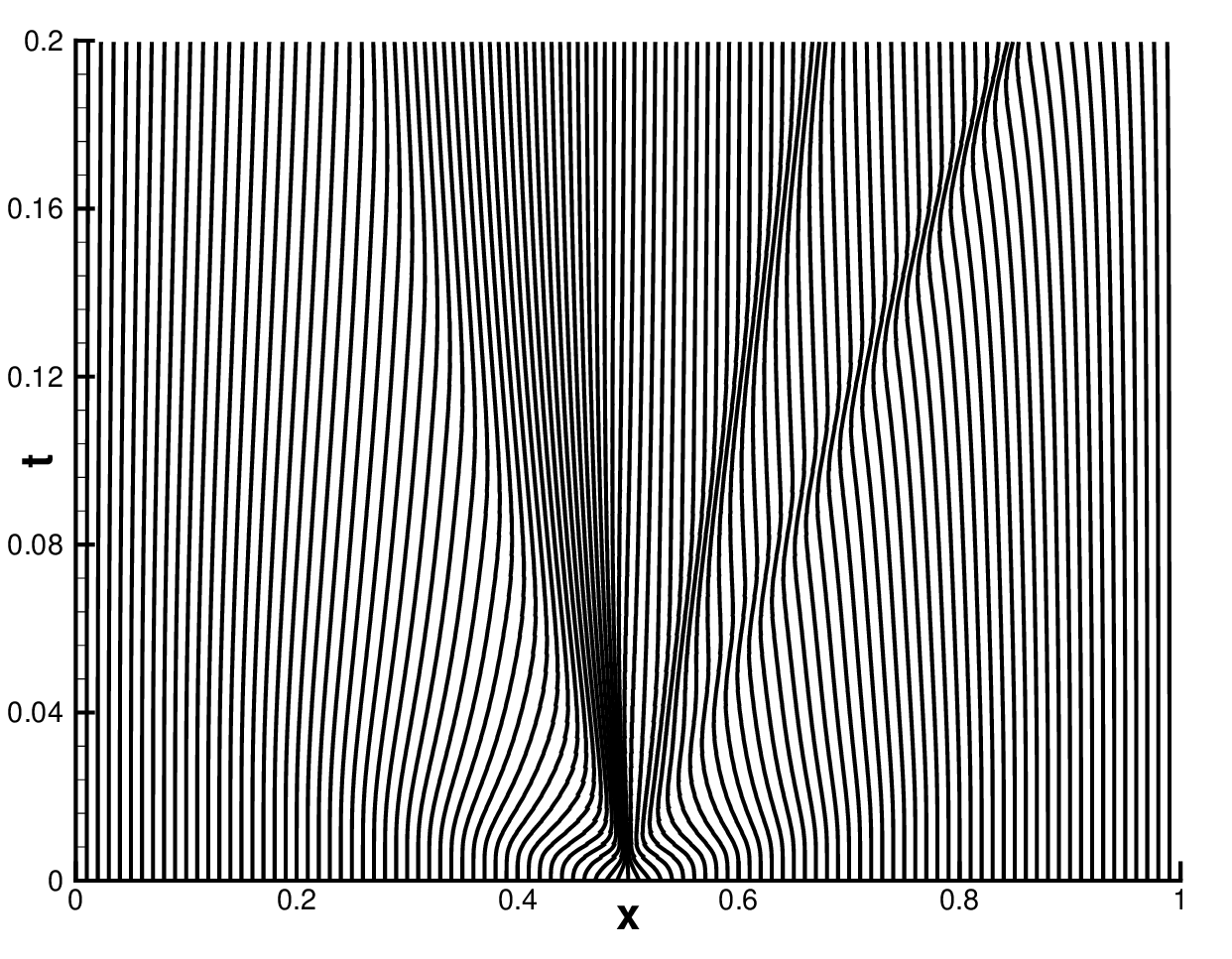}
\caption{\label{sod-mesh} One-dimensional Riemann  problem: the history of mesh evolution for Lagrangian velocity (left) and adaptation velocity (right) for Sod problem.}
\end{figure}

\begin{figure}[!htb]
  \centering
  \includegraphics[width=0.475\textwidth]{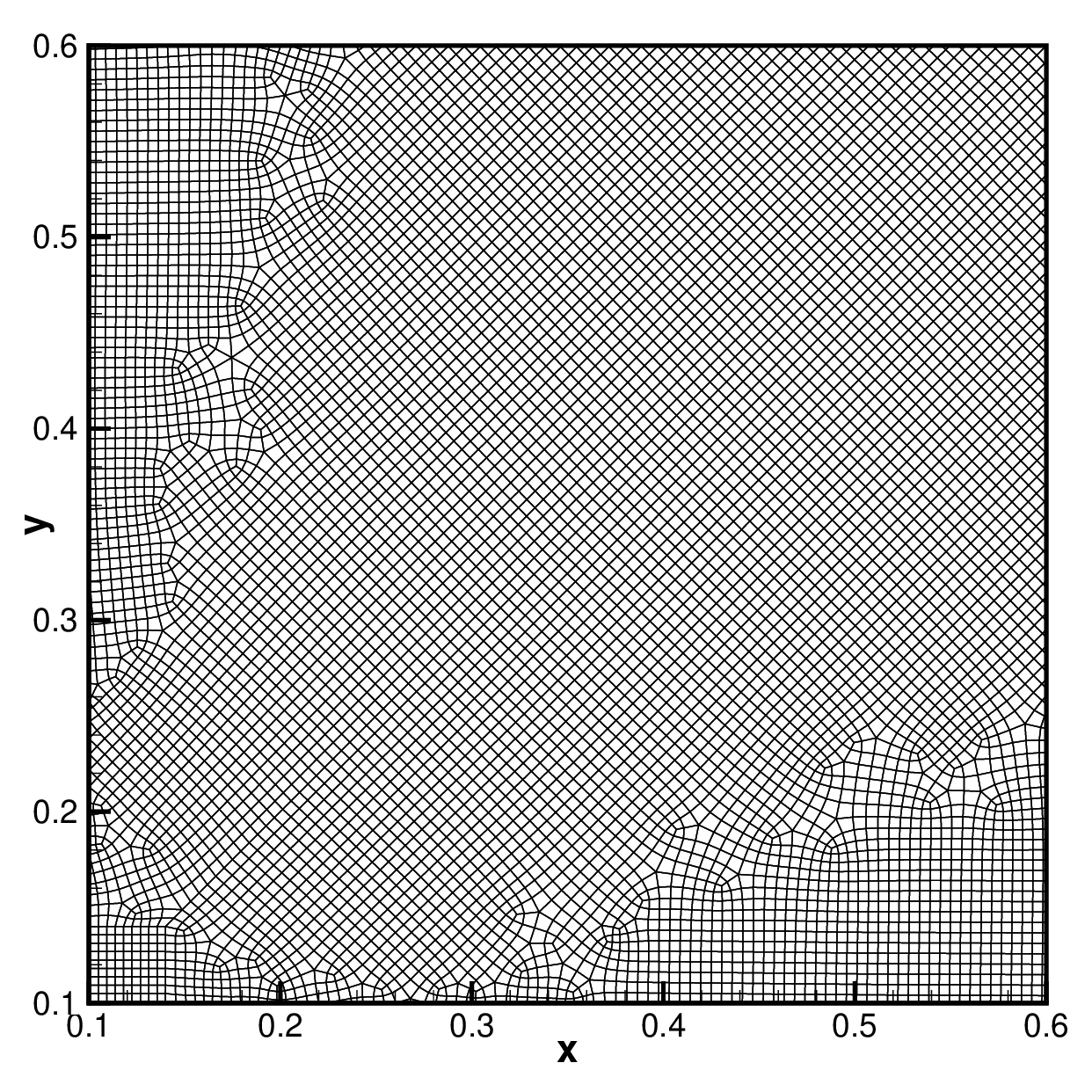} 
  \includegraphics[width=0.475\textwidth]{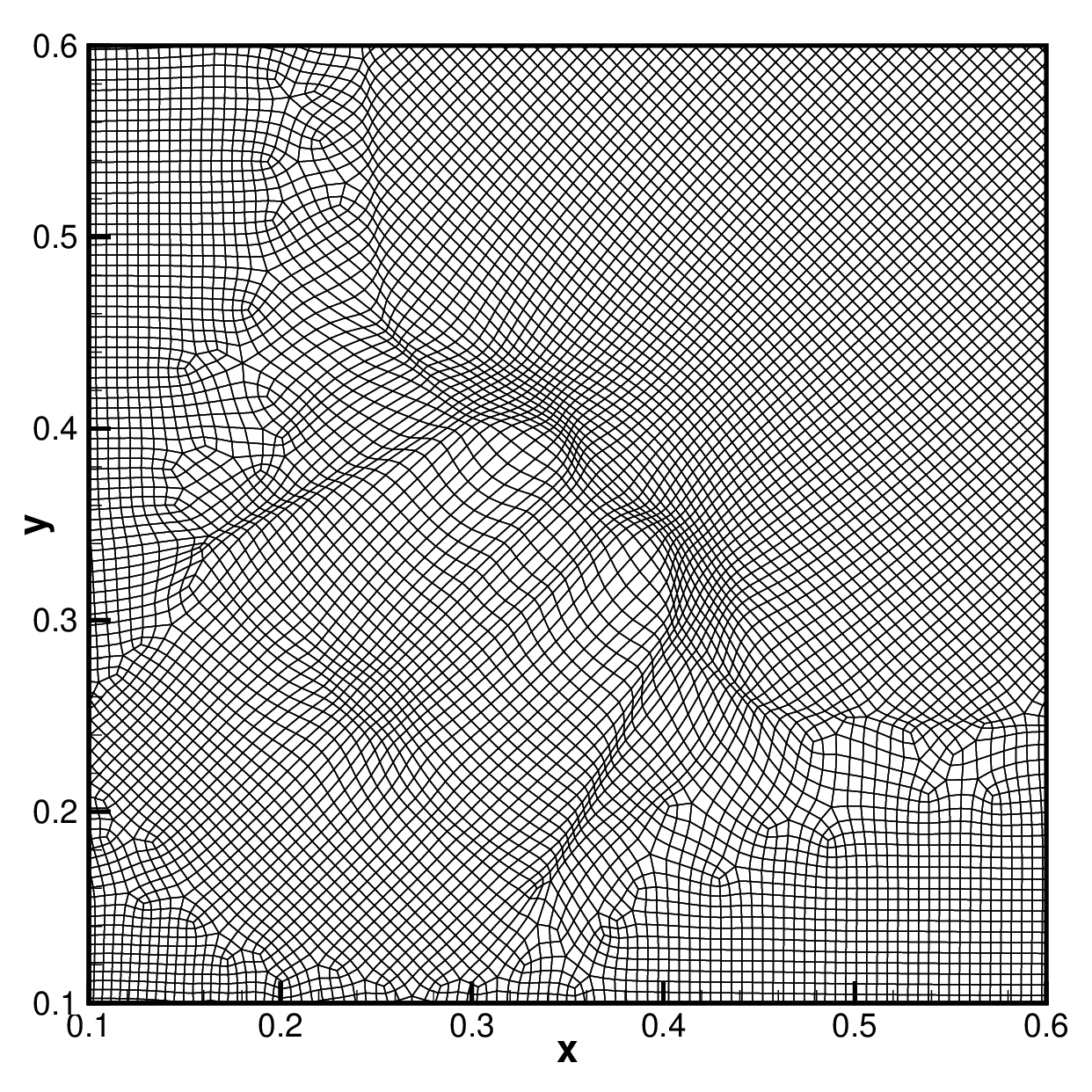}
  \caption{\label{Rie4-100} Two-dimensional Riemann problem:  the local enlarged mesh distribution at $t=0$ (left) and $t=0.4$ (right) using adaptive mesh with mesh size $h=1/200$. }
  \centering
  \includegraphics[width=0.475\textwidth]{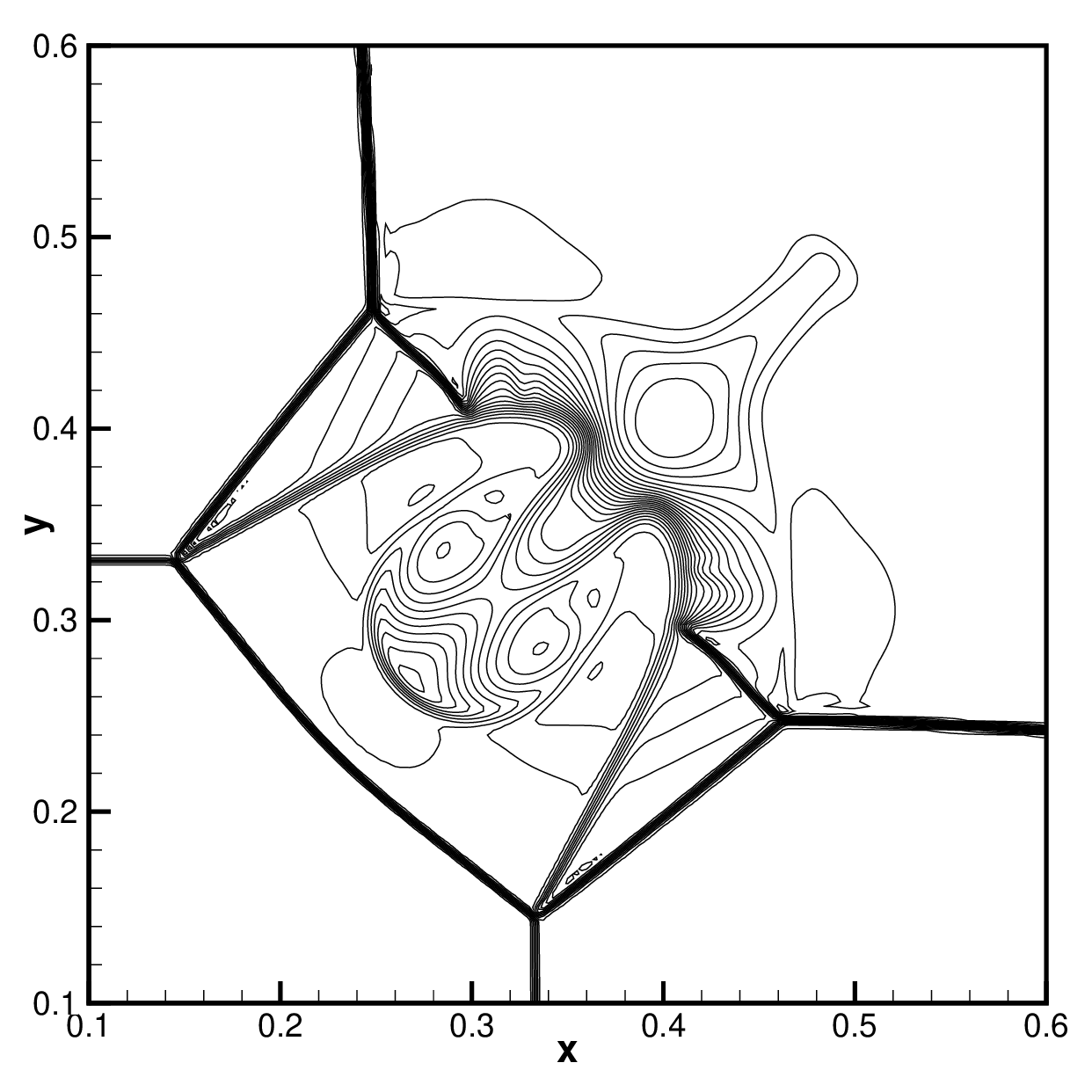} 
  \includegraphics[width=0.475\textwidth]{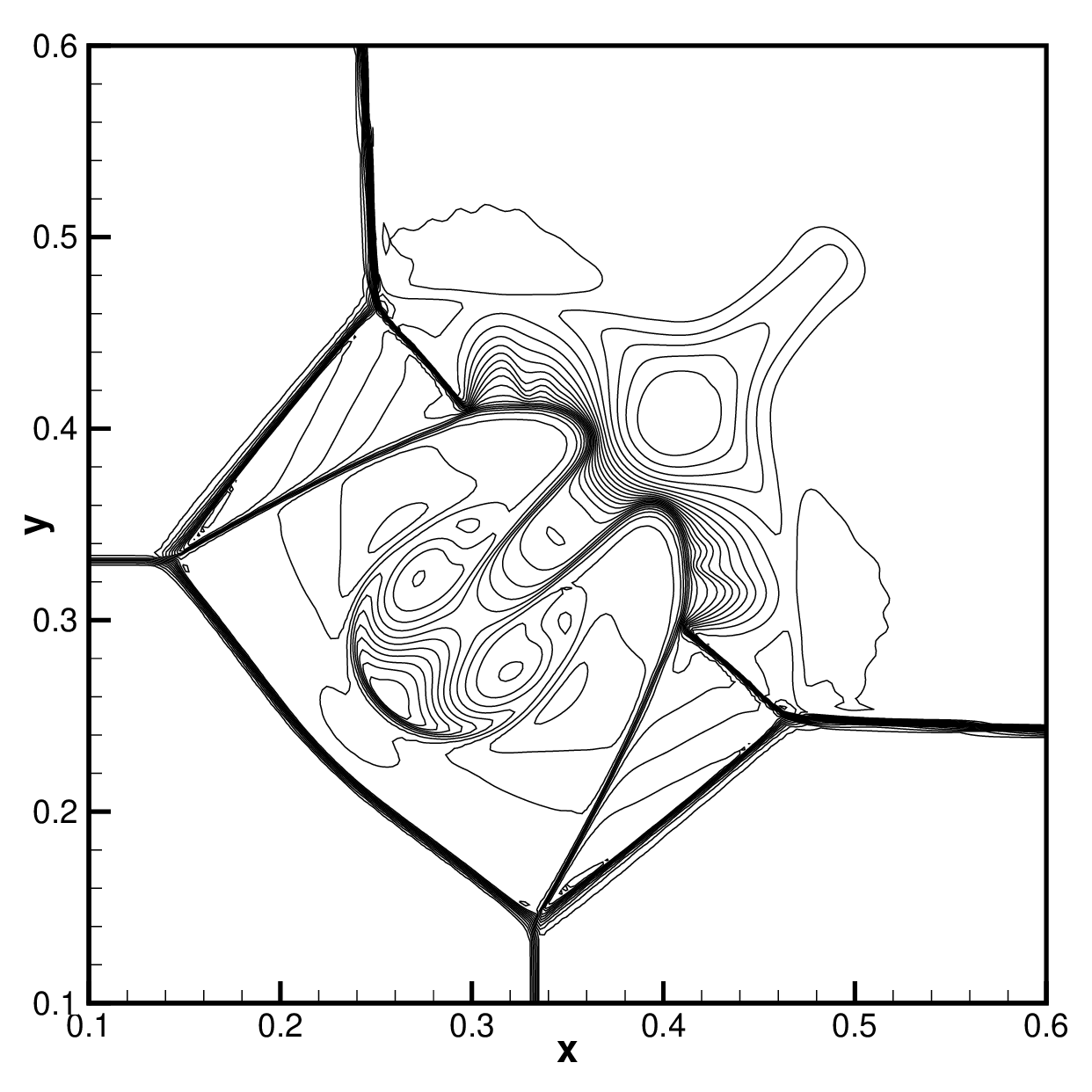}
  \caption{\label{Rie4-400} Two-dimensional Riemann problem: the local enlarged density distribution using stationary mesh (left) and adaptive mesh (right) with mesh size $h=1/400$. }
\end{figure}

\subsection{One-dimensional Riemann problem}
In this case, the Sod problem is considered for one-dimensional Riemann problem, and the initial condition is given by
\begin{equation*}
(\rho, U,  p)=\begin{cases}
    (1, 0,  1),  \ \ \ \ &  0\leq x<0.5,\\
    (0.125, 0,  0.1),    & 0.5<x\leq1.
  \end{cases}
\end{equation*}
The computational domain is $[0,1]\times[0,0.1]$, and the uniform mesh with cell size $h=1/100$ is adopted. 
The non-reflective conditions are applied at all boundaries. 
The mesh velocity can be given by the variational approach \cite{Adaptive-Tang_HZ-2002} for local mesh adaptation, 
and the Lagrangian nodal solver \cite{Lagrangian-c1} to track the material interface. 
The smooth procedure is used every 20 steps for the Lagrangian velocity and adaptive velocity. 
The density distribution at $t=0.2$ is presented in Fig.\ref{sod-density},  
The history of mesh evolution for Lagrangian velocity and adaptive velocity is shown in Fig.\ref{sod-mesh}. 
The numerical solutions agree well with the exact solutions. 
Due to the mesh adaptation procedure, the discontinuities are well captured by the current scheme.

\subsection{Two-dimensional Riemann problem}
In this case, an example of two-dimensional Riemann problem for Euler equations are presented \cite{Case-Lax}. 
The computational domain is $[0,1]\times[0,1]$ and non-reflective condition is used for all boundaries. 
The adaptation velocity is chosen as mesh velocity, and the parameter $\alpha$ in the monitor function takes $0.1$. 
The initial condition for this problem is given by
\begin{align*}
  (\rho,U,V,p)&= 
  \left\{
  \begin{aligned}
    &(1.5,0,0,1.5),              & x>0.5,y>0.5, \\
    &(0.5323,1.206,0,0.3),       & x<0.5,y>0.5, \\
    &(0.138,1.206,1.206,0.029),  & x<0.5,y<0.5, \\
    &(0.5323,0,1.206,0.3),       & x>0.5,y<0.5.
  \end{aligned} 
  \right.
\end{align*}
In this case, four initial shock waves interact with each other and result in a much more complicated pattern. 
The initial shock wave $S^-_{23}$ bifurcates at the trip point into a reflected shock wave, a Mach stem, and a slip line. 
The reflected shock wave interacts with the shock wave $S^-_{12}$, which produces a new shock. 
In the computation, the unstructured quadrilateral mesh is used. 
The local enlargement of the mesh at $t=0$ and $t=0.4$ are given in Fig.\ref{Rie4-100}, 
where the mesh size $h=1/200$, and the variational approach \cite{Adaptive-Tang_HZ-2002} is used for the local mesh adaptation. 
It can be observed that the mesh adaptively concentrates in zones with high density gradient. 
The local enlargement of density distribution with cell size $h=1/400$ is given in Fig.\ref{Rie4-400}, 
where the small-scale flow structures are well captured by the current scheme. 
As reference, the numerical results with stationary mesh is given in Fig.\ref{Rie4-400} as well, 
where the discontinuities are smeared much more compared with the case of adaptive mesh.

\begin{figure}[!htb]
\centering
\includegraphics[width=0.95\textwidth]{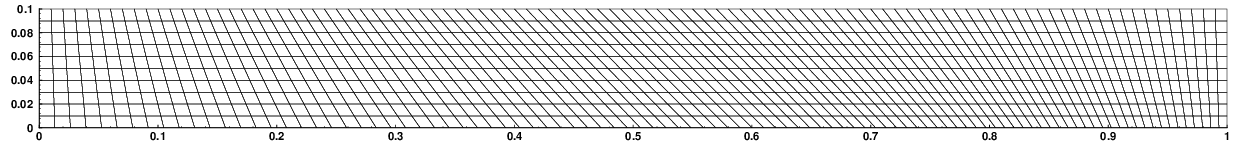} 
\includegraphics[width=0.95\textwidth]{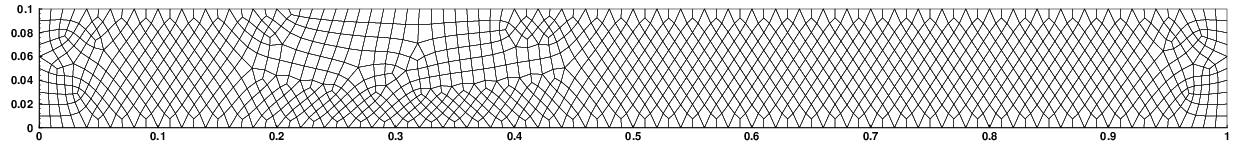} 
\caption{\label{saltz-mesh0} Saltzman problem: the initial mesh distribution for structured mesh (top) and unstructured mesh (bottom). }
\end{figure}

\begin{figure}[!htb]
\centering
\includegraphics[width=0.475\textwidth]{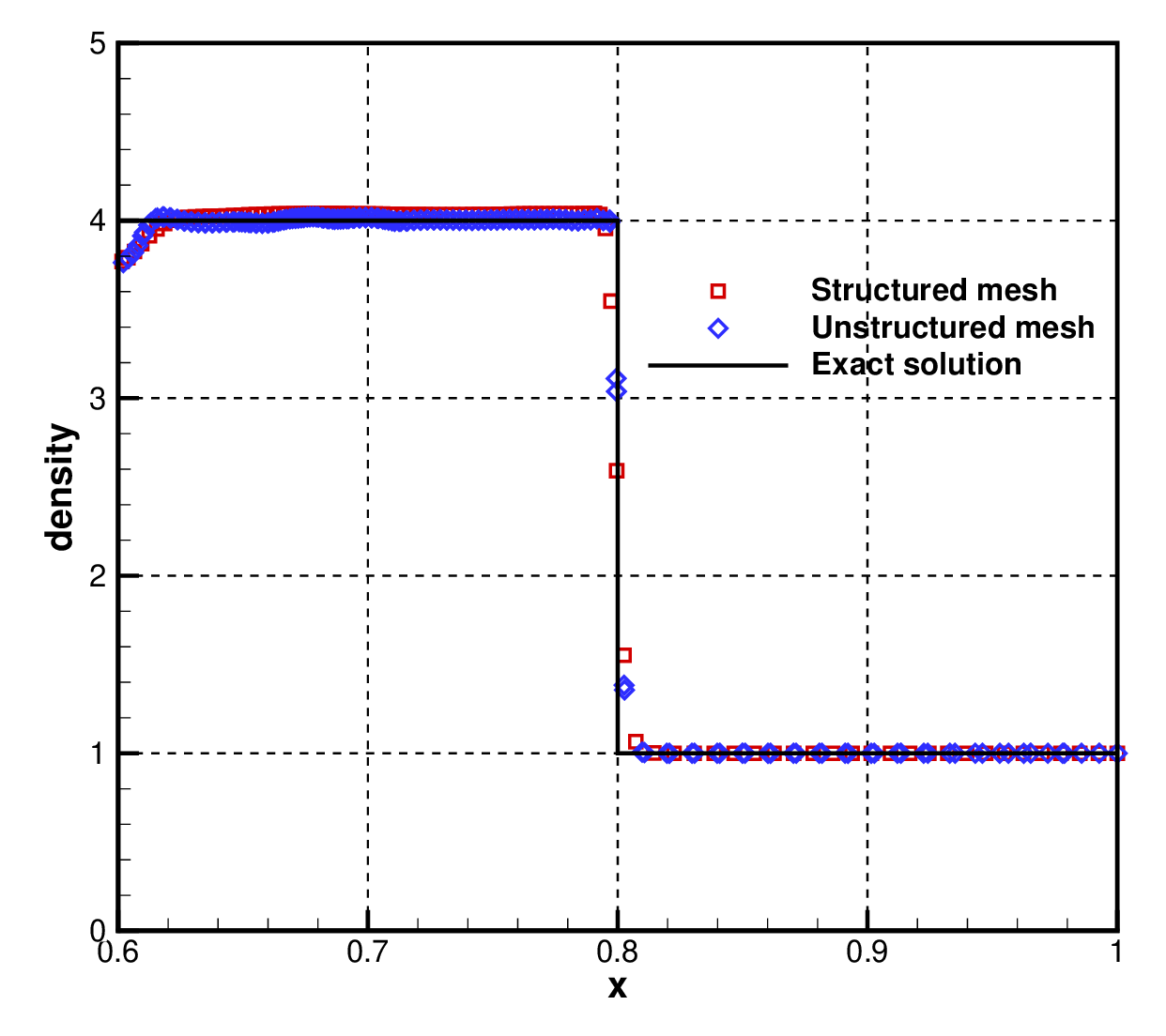} 
\includegraphics[width=0.475\textwidth]{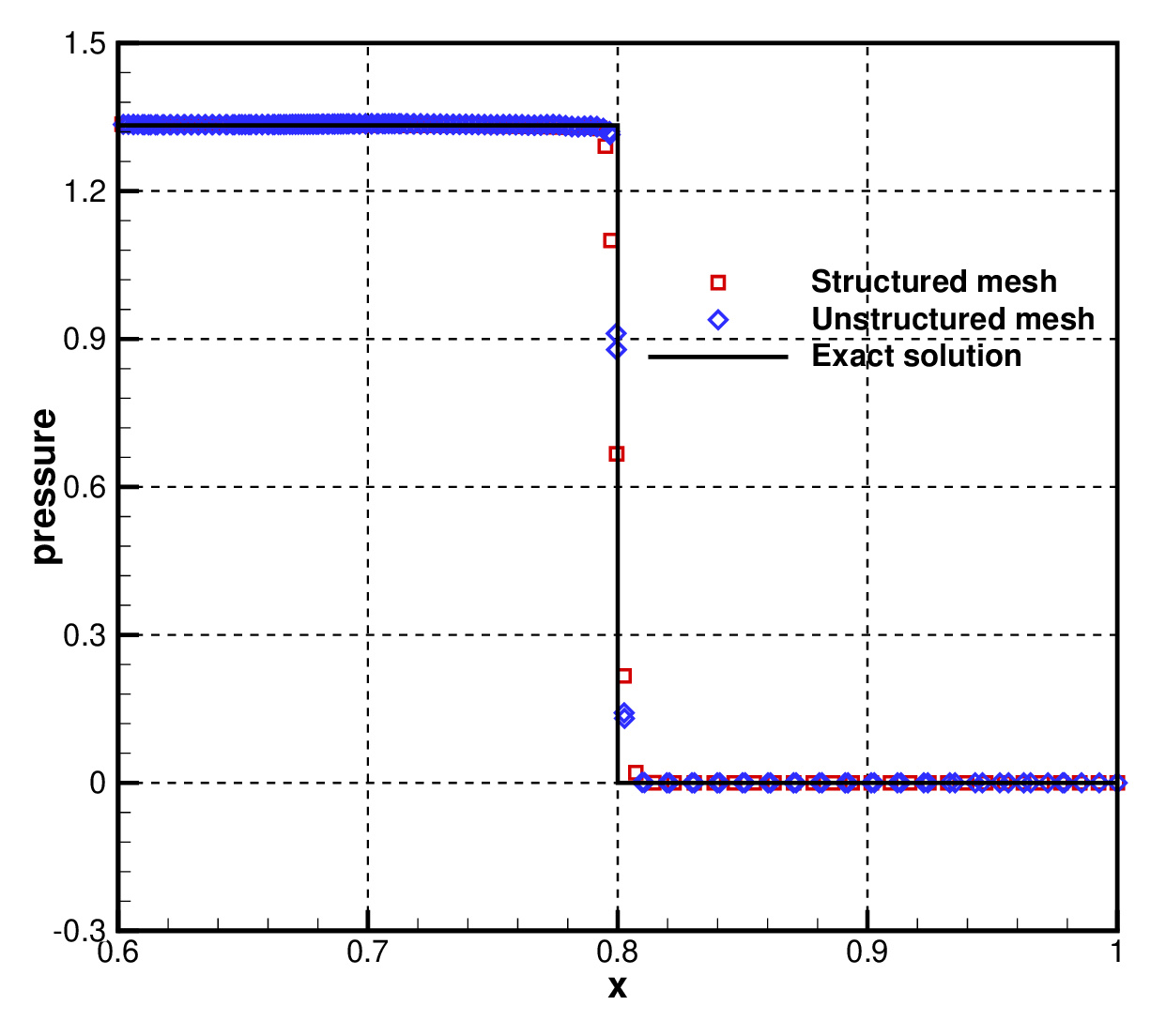}
\caption{\label{saltz-data} Saltzman problem: the density and pressure profiles at $y=0.02$ for structured mesh (left) and unstructured mesh (right)  at $t=0.6$.}
\end{figure}

\begin{figure}[!htb]
\centering
\includegraphics[width=0.75\textwidth]{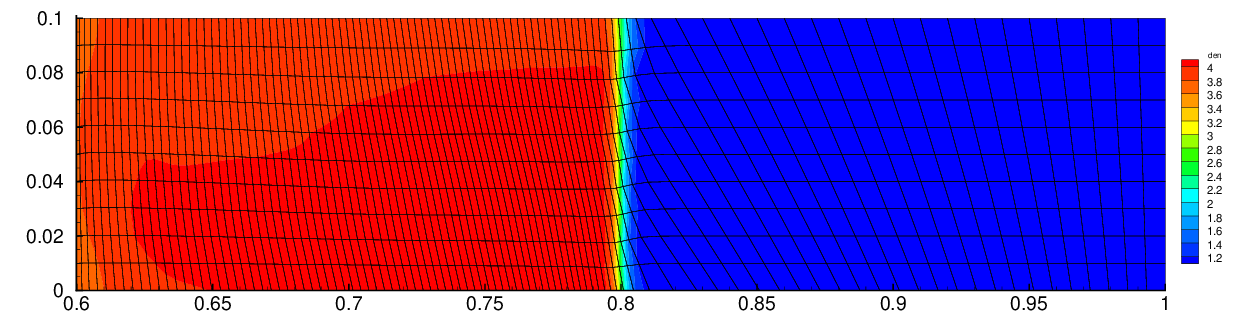} \\
\includegraphics[width=0.75\textwidth]{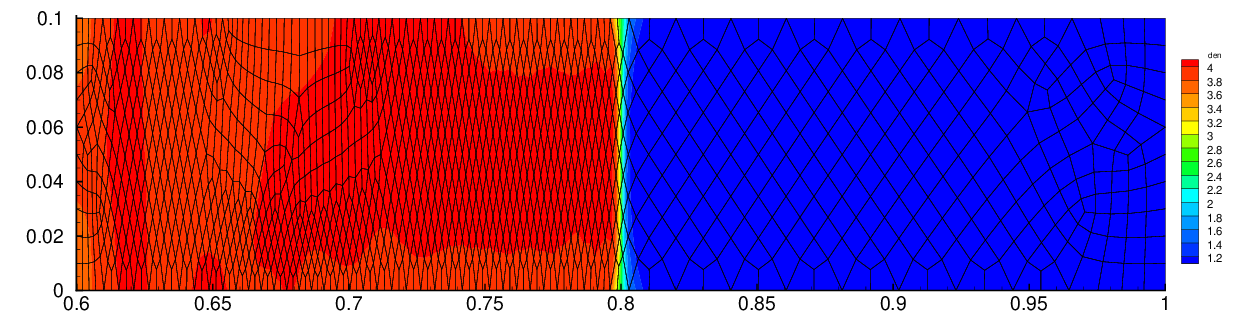} \\
\caption{\label{saltz-mesh4} Saltzman problem: the mesh and density distribution for structured mesh (top) and unstructured mesh (bottom) at $t=0.6$. }
\end{figure}

\subsection{Saltzmann problem}
This is a benchmark test case for the Lagrangian and ALE codes, which tests the ability to 
capture shock propagation with a systematically distorted mesh \cite{Case-Saltzman-1,Case-Saltzman-2}. 
The computational domain is $[0, 1] \times [0, 0.1]$, and the initial distorted structured and unstructured meshes are given in Fig.\ref{saltz-mesh0}.
An ideal monatomic gas with $\rho = 1, e = 10^{-4}, \gamma = 5/3$ is filled in the box. 
The left-hand side wall acts as a piston with a constant velocity $U_p$ = 1, and other boundaries are reflective walls. 
As the gas is compressed, a strong shock wave is generated from the left end. 
The Lagrangian nodal solver \cite{Lagrangian-c1} is chosen for mesh velocity, and the smoothing process is applied every 20 iteration steps. 
At $t=0.6$, the shock is expected to be located at $x=0.8$, and the post shock solutions are $p = 4$ and $p = 1.333$. 
The density and pressure distributions at $y=0.02$ for structured and unstructured meshes are given in Fig.\ref{saltz-data}, and the numerical results agree well with the exact solutions. 
The mesh and density profiles for structured and unstructured meshes are given in Fig.\ref{saltz-mesh4} at $t = 0.6$. 
The meshes are distorted severely in the compression process, especially for the unstructured mesh. 
The case validates the robustness of current scheme with the unstructured mesh.

\begin{figure}[!htb]
\centering
\includegraphics[width=0.475\textwidth]{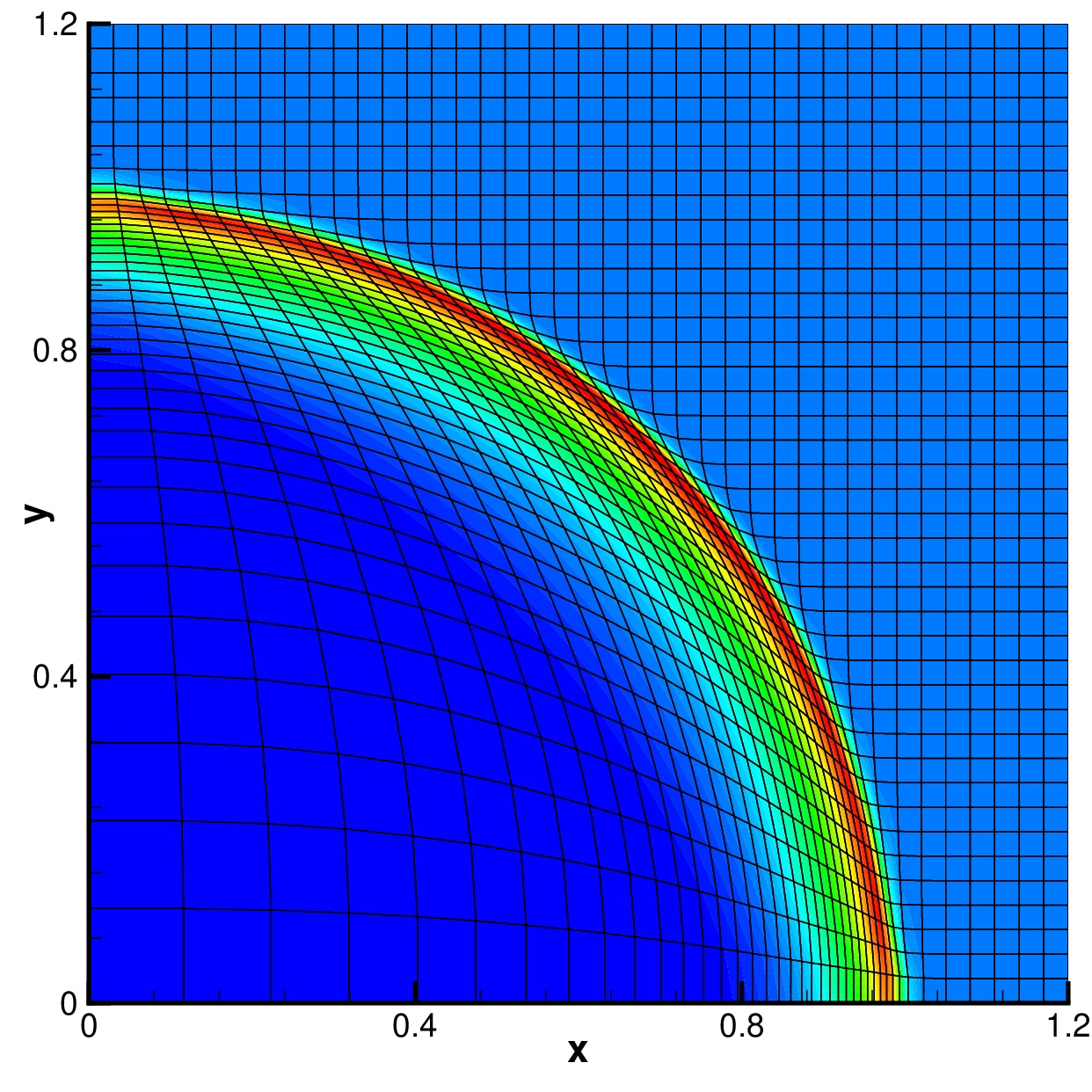}
\caption{\label{sedov-mesh} Sedov problem: the density and mesh distributions at $t=1$ with mesh size $h=1/40$. }
\centering
\includegraphics[width=0.475\textwidth]{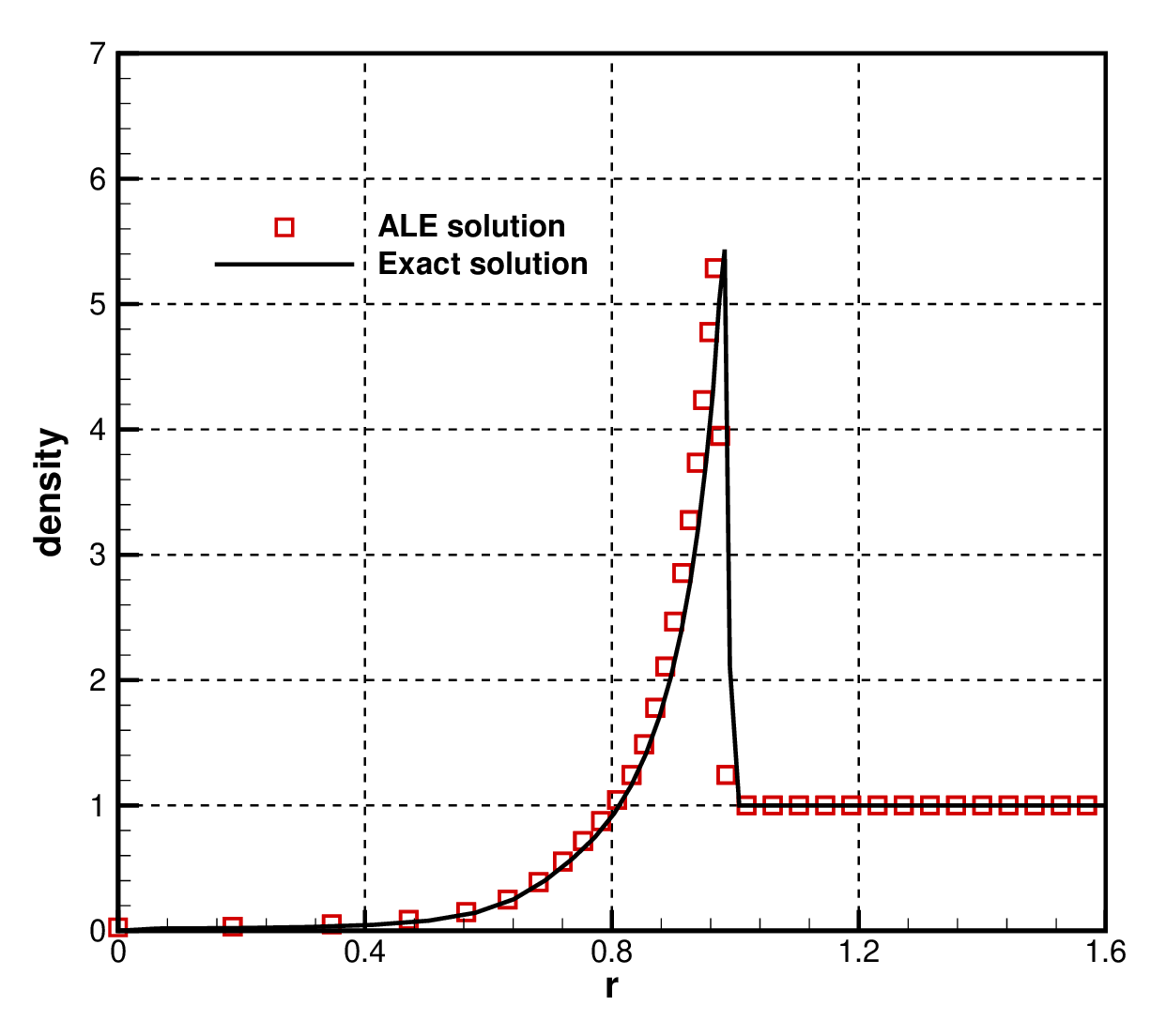} 
\includegraphics[width=0.475\textwidth]{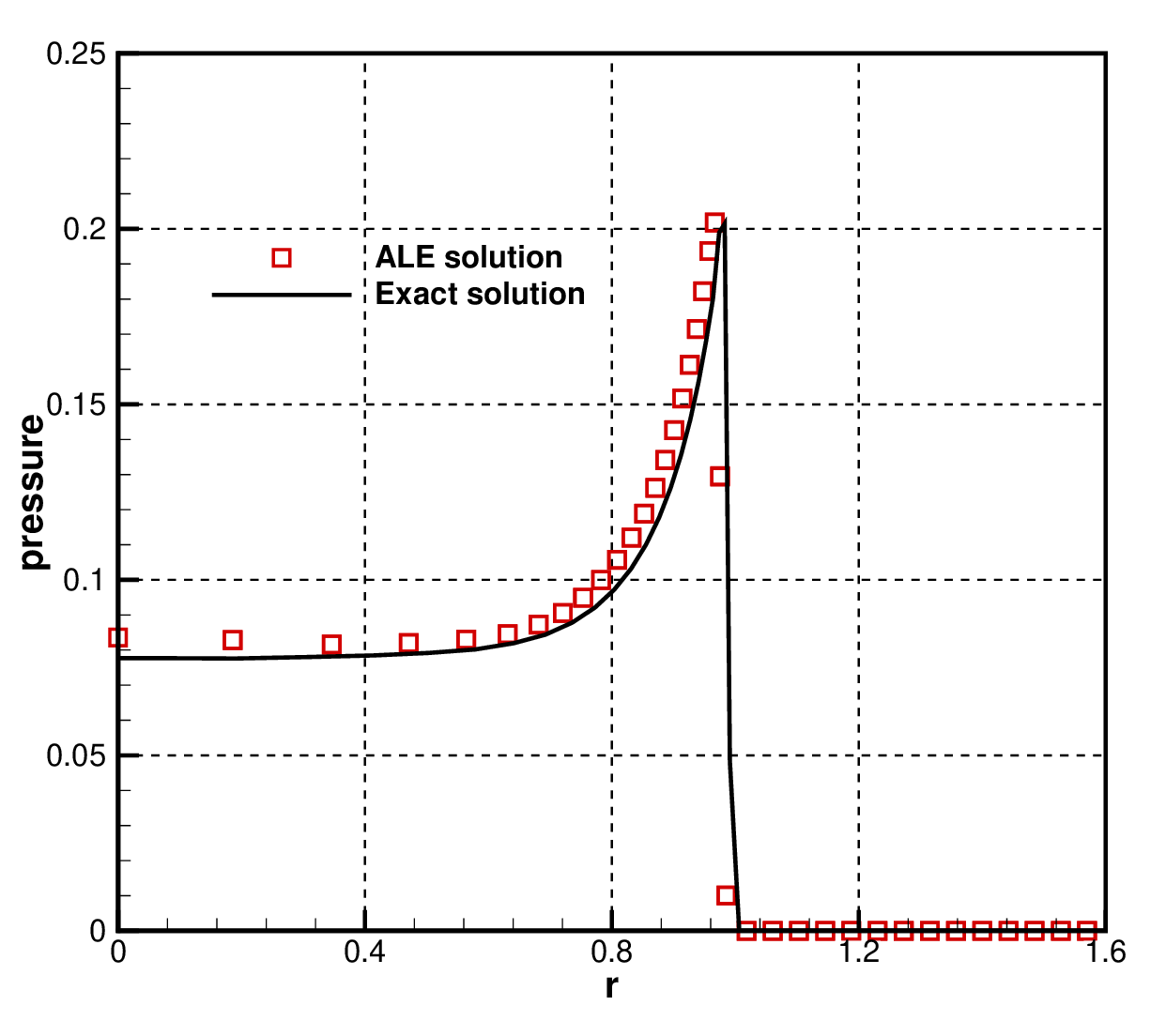}
\caption{\label{sedov-data} Sedov problem: the density and pressure profiles along the diagonal line with mesh size $h=1/40$ at $t=1$.}
\end{figure}

\subsection{Sedov problem}
The Sedov problem is a two-dimensional explosion problem to model blast wave from an energy deposited singular point \cite{Case-Sedov}, which is a standard benchmark problem for the Lagrangian method. 
The fluid is modeled by the ideal gas EOS with $\gamma=1.4$. The initial density has a uniform unit distribution, and the pressure is $10^{-6}$ everywhere, except in the cell containing the origin. 
For the cell containing the origin, the pressure is defined as $p=(\gamma-1)\varepsilon_0/V$, where $\varepsilon_0=0.244816$ is the total amount of released energy and $V$ is the cell volume.  
According to the analysis of self-similar solution \cite{Case-Sedov}, the solution consists of a diverging infinite strength shock wave whose front is located at radius $r=1$ at $t=1$. 
The computational domain is $[0,1.2]\times[0,1.2]$ and the uniform mesh with cell size $h=1/40$ is adopted. 
The reflective condition is used for the left and bottom boundaries, and non-reflective condition is used for the right and top boundaries. 
The Lagrangian nodal solver \cite{Lagrangian-c1} is chosen for mesh velocity, and the smoothing process is applied every 20 iteration steps. 
The density and mesh distribution at $t=1$ is given in Fig.\ref{sedov-mesh}.  
The density and pressure profile along the diagonal line $t=1$ are given in Fig.\ref{sedov-data}, and the numerical results agree well with the exact solutions.
The robustness of current scheme is validated by the strong explosive wave.

\begin{figure}[!htb]
\centering
\includegraphics[width=0.45\textwidth]{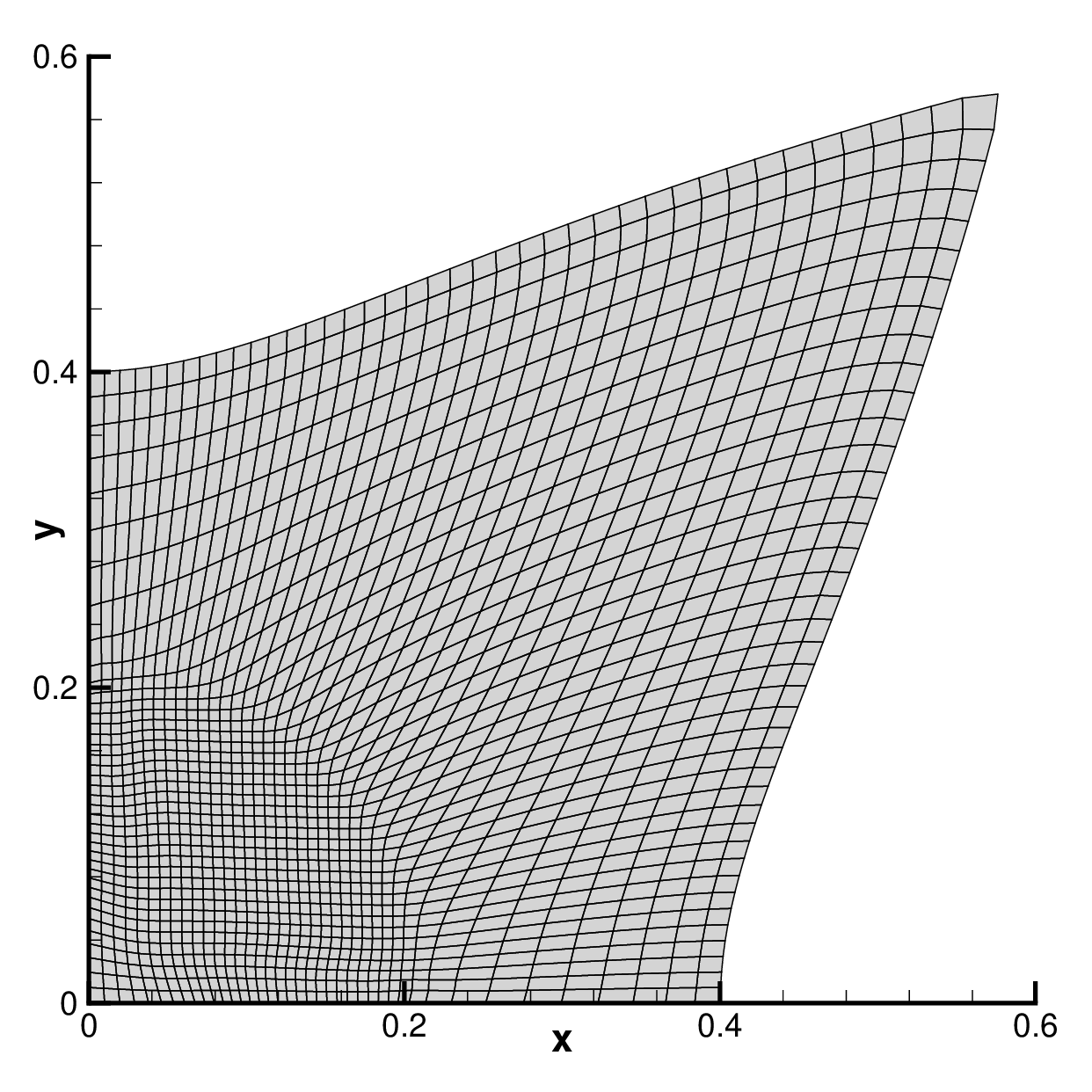}
\caption{\label{noh-flow-1} Noh problem: the mesh distribution with mesh size $h=1/40$ at $t=0.6$. }
\includegraphics[width=0.45\textwidth]{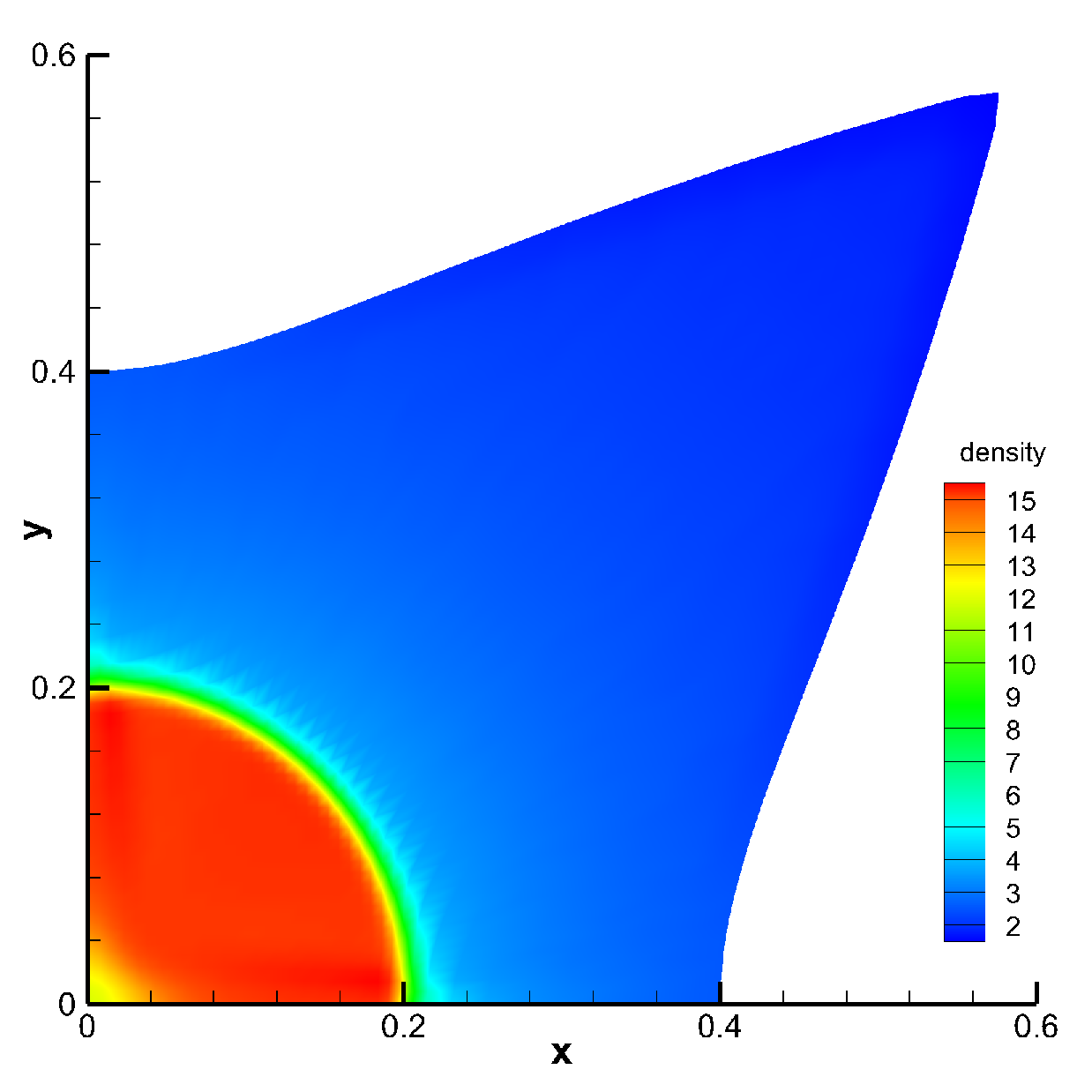}
\includegraphics[width=0.45\textwidth]{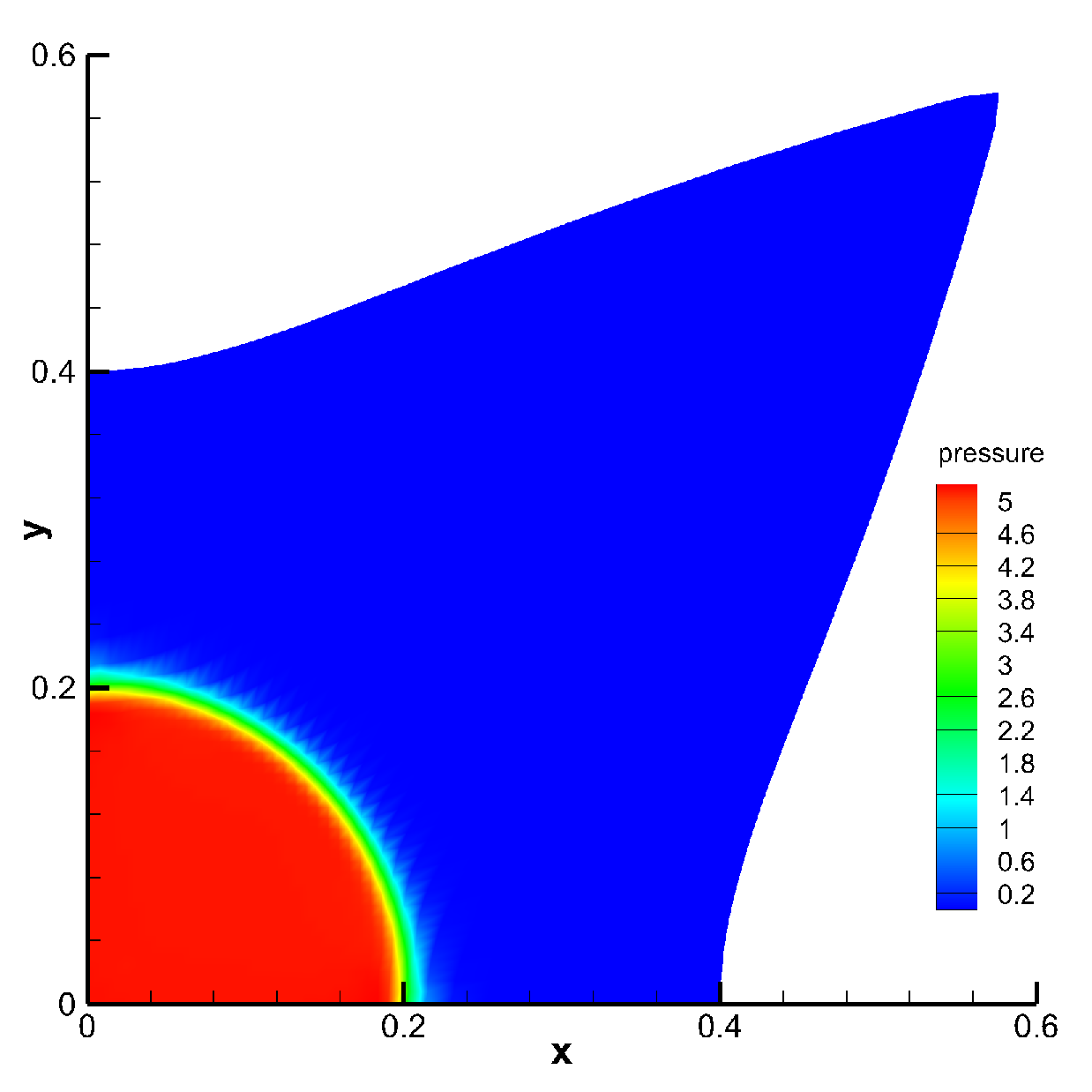}
\caption{\label{noh-flow-2} Noh problem: the density distribution (left) and  pressure distribution (right) with mesh size $h=1/40$ at $t=0.6$.}
\end{figure}   

\begin{figure}[!htb]
\centering 
\includegraphics[width=0.475\textwidth]{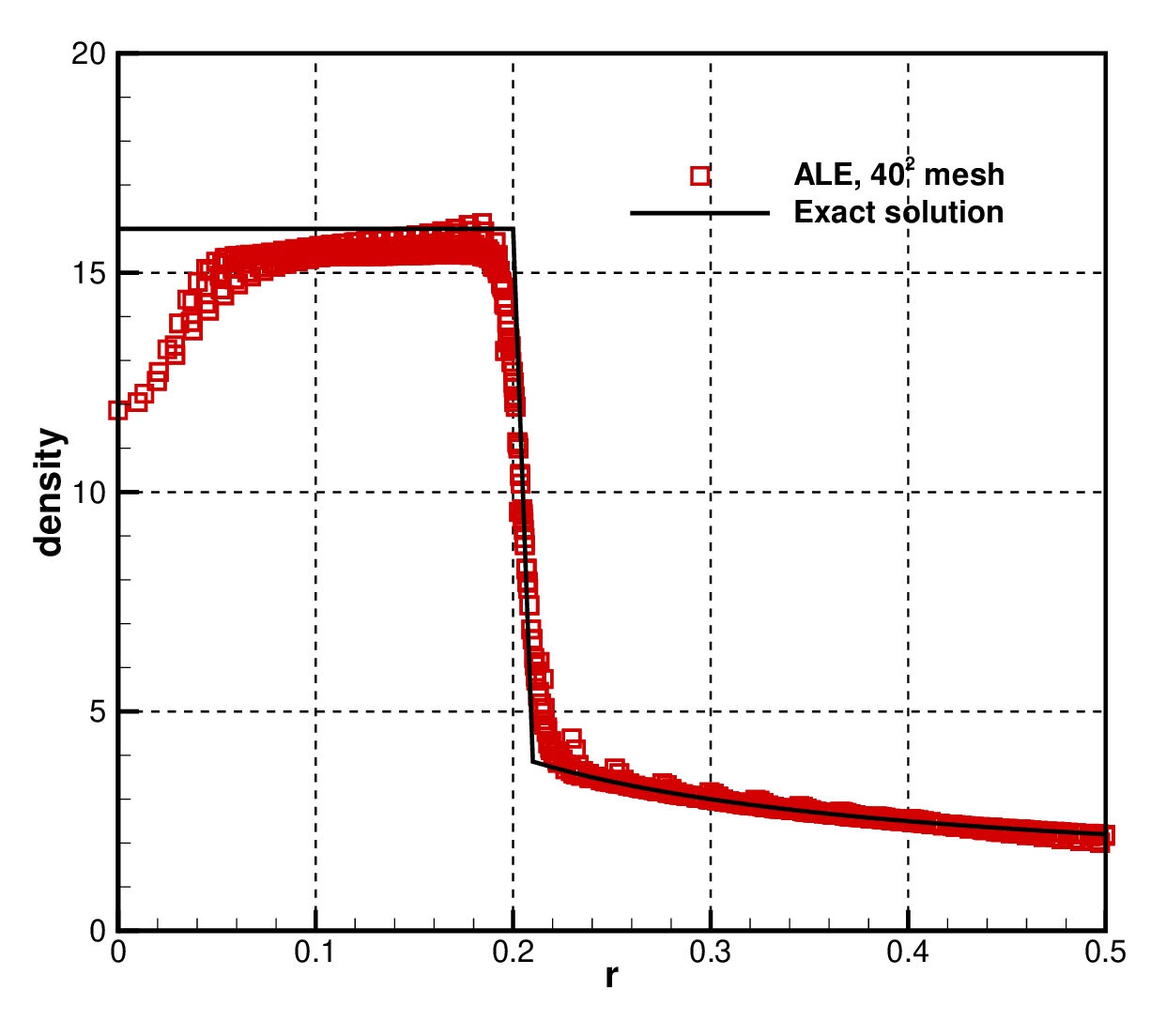} 
\includegraphics[width=0.475\textwidth]{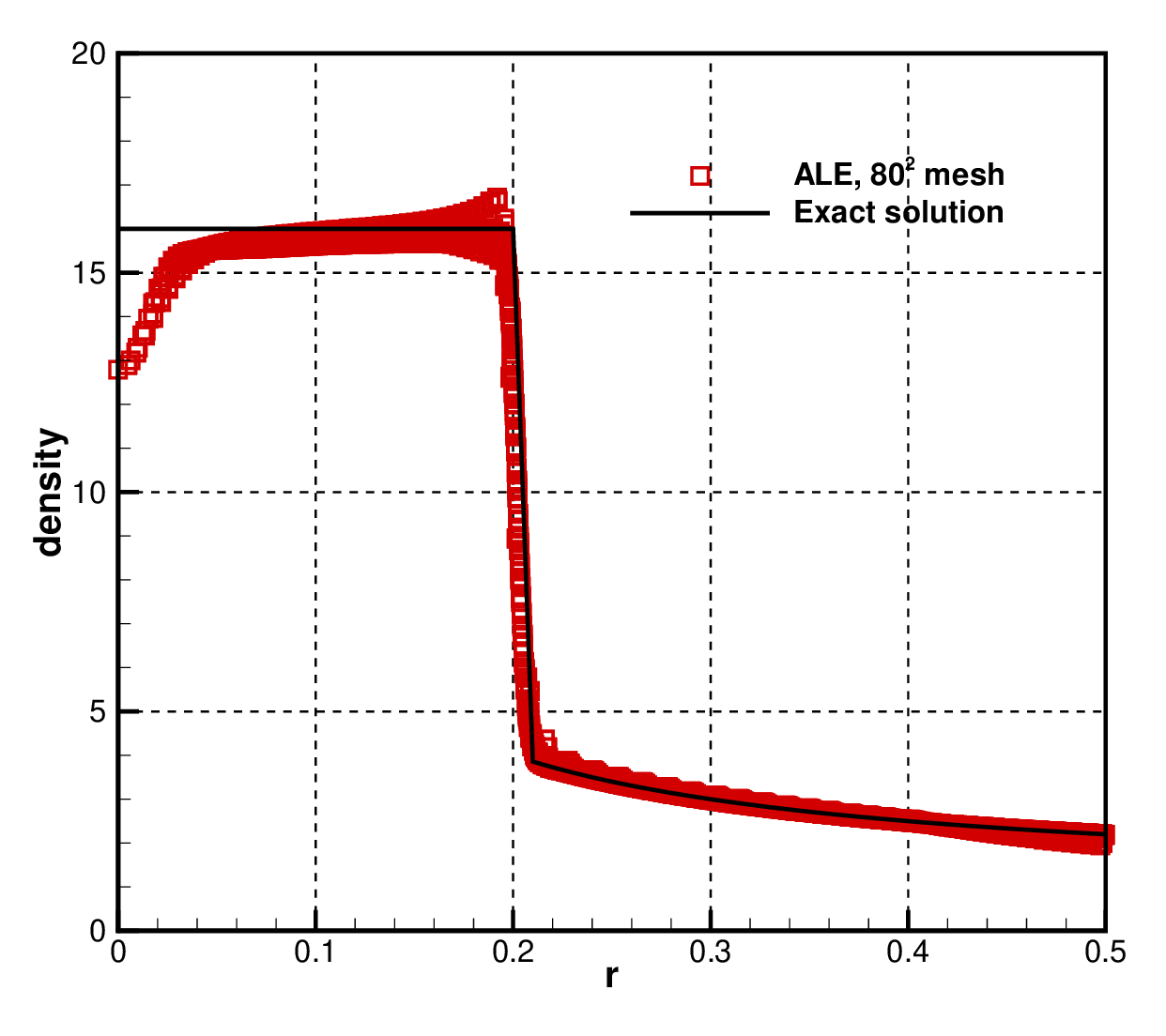} 
\caption{\label{noh-data}  Noh problem: the density distributions in all cells with mesh size $h=1/40$ (left) and mesh size $h=1/80$ (right) at $t=0.6$. }
\end{figure}

\subsection{Noh Problem}
The Noh problem is a typical test case for verifying ALE code \cite{Case-Noh}. The initial conditions are set as follows
\begin{align*}
 (\rho, U,V, e)= (1, -x/\sqrt{x^2+y^2}, -y/\sqrt{x^2+y^2}, 1\times 10^{-4}), 
\end{align*}
where $e = p/(\gamma-1)$ and $\gamma=5/3$. An axially symmetric shock is generated at the origin and propagates at a constant speed. 
At time $t = 0.6$, the shock is located at $r = 0.2$, and the exact solution of density is given by
\begin{align*}
\rho=\begin{cases}
16, &r < 0.2,\\
1+t/r, &r>0.2. 
\end{cases}
\end{align*}
In the computation, the two-dimensional domain $[0, 1] \times [0, 1]$ is considered. 
The asymmetric boundary condition is enforced at the lines $x = 0$ and $y = 0$, 
while a non-reflective boundary condition is used at other boundaries. 
A uniform unstructured mesh with mesh size $h = 1/40$ and $h = 1/80$ is adopted. 
The Lagrangian nodal solver \cite{Lagrangian-c1} is used for mesh moving procedure, and a smoothing process is applied every 20 iteration steps. 
The final mesh distribution is shown in Fig.\ref{noh-flow-1}, where the mesh is concentrated nicely in the post-shock region. 
The density and pressure distributions at time $t = 0.6$ is shown in Fig.\ref{noh-flow-2}, and the density 
distributions with respect to $r$ in all cells is presented in Fig.\ref{noh-data}, which agree well with the exact solution.

\begin{figure}[!h]
\centering
\includegraphics[width=0.475\textwidth]{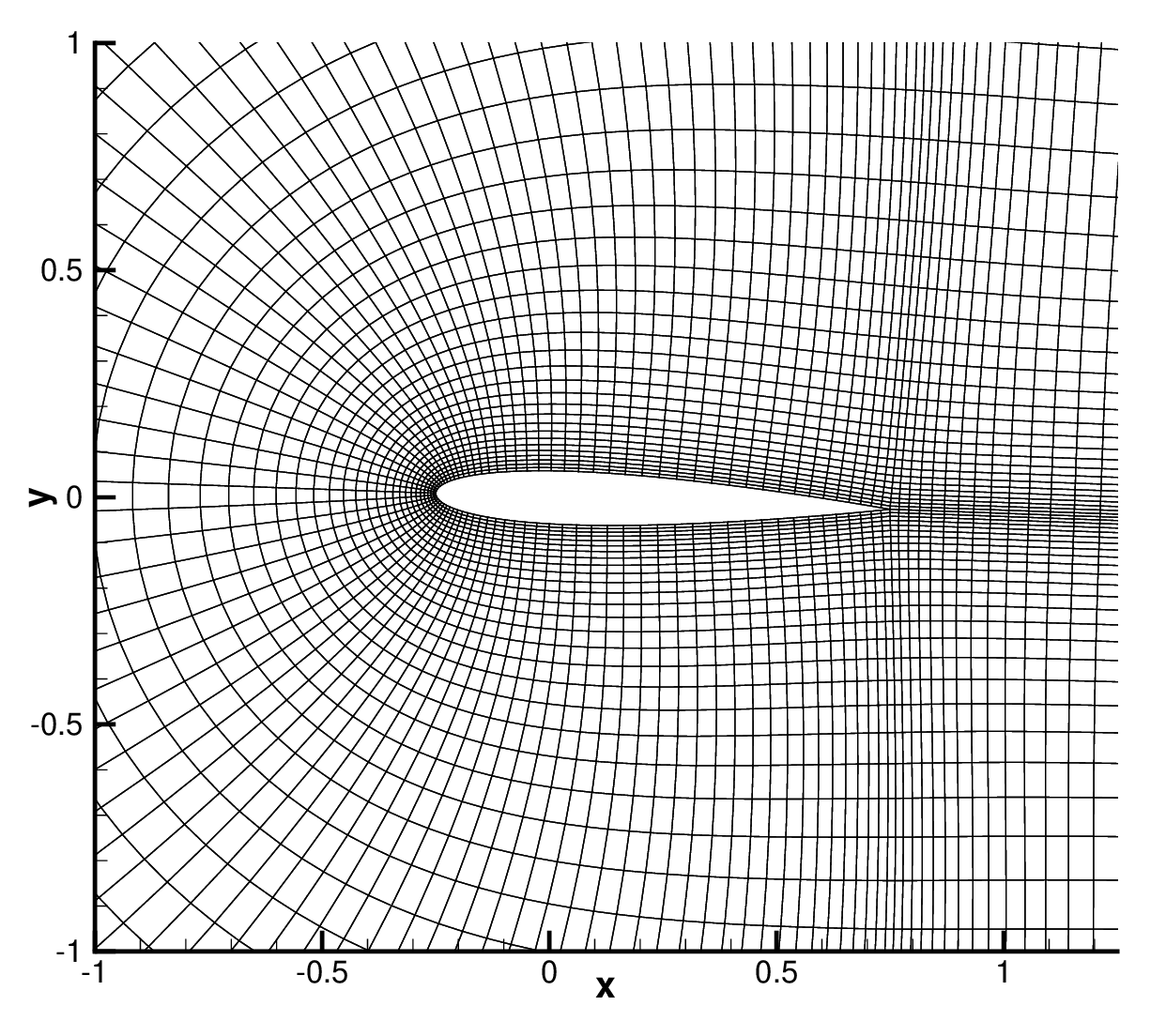}
\includegraphics[width=0.475\textwidth]{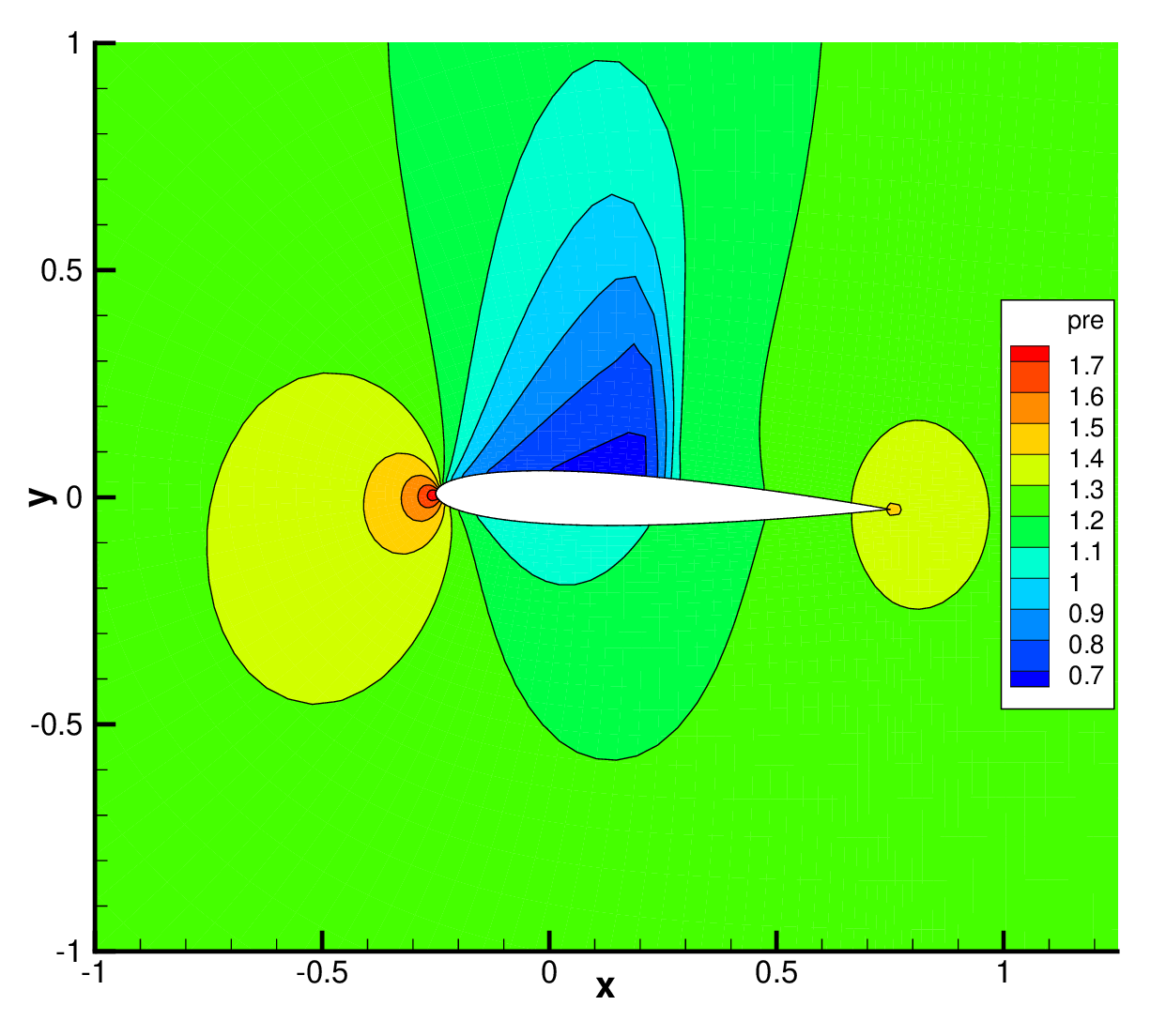}
\caption{\label{airfoli-field} Flow over NACA 0012 airfoil: the mesh and pressure distribution at the angle of attack $2.01^\circ$.}
\end{figure}

\subsection{Flow over NACA 0012 Airfoil}
\begin{figure}[!h]
\centering
\includegraphics[width=0.45\textwidth]{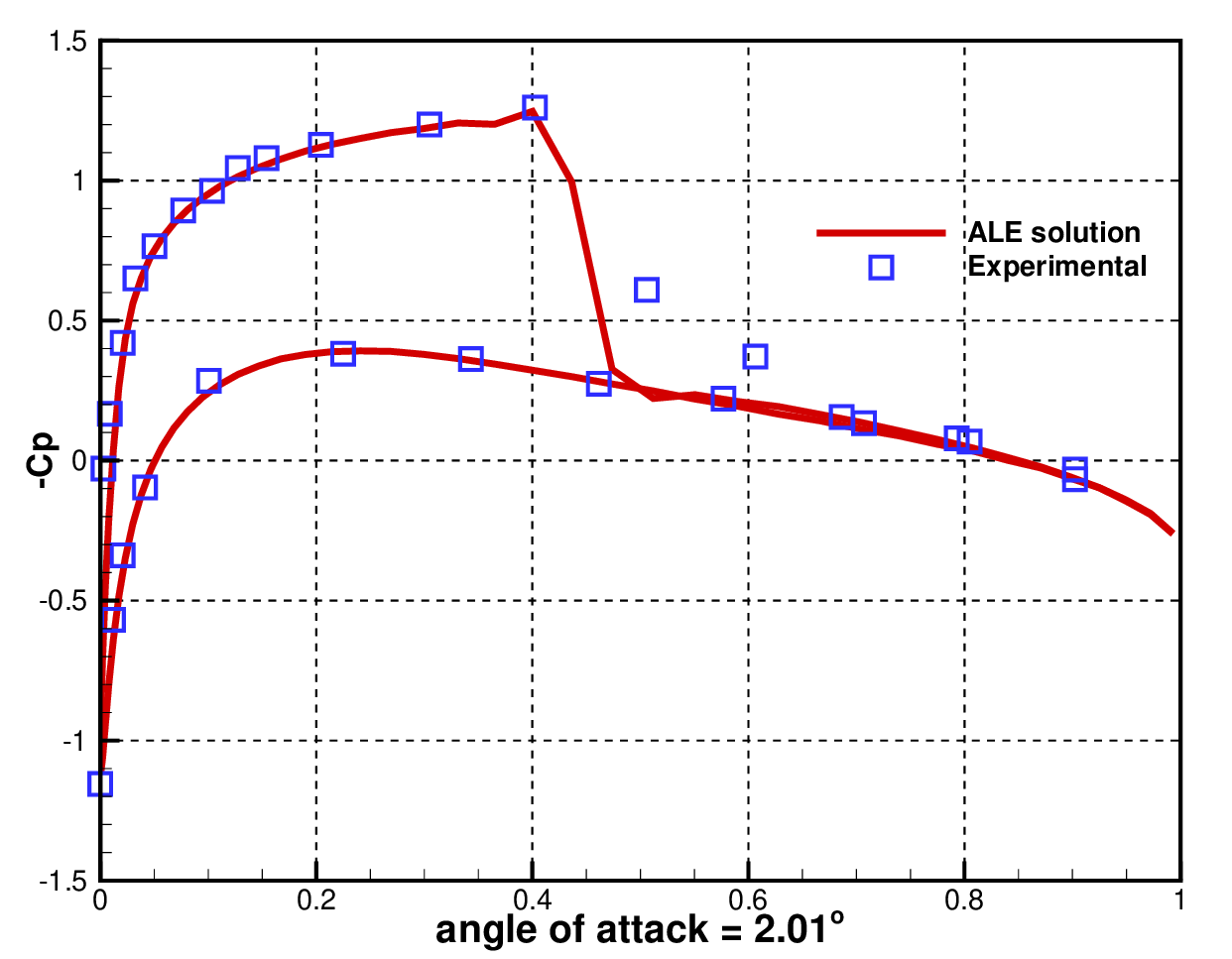}
\includegraphics[width=0.45\textwidth]{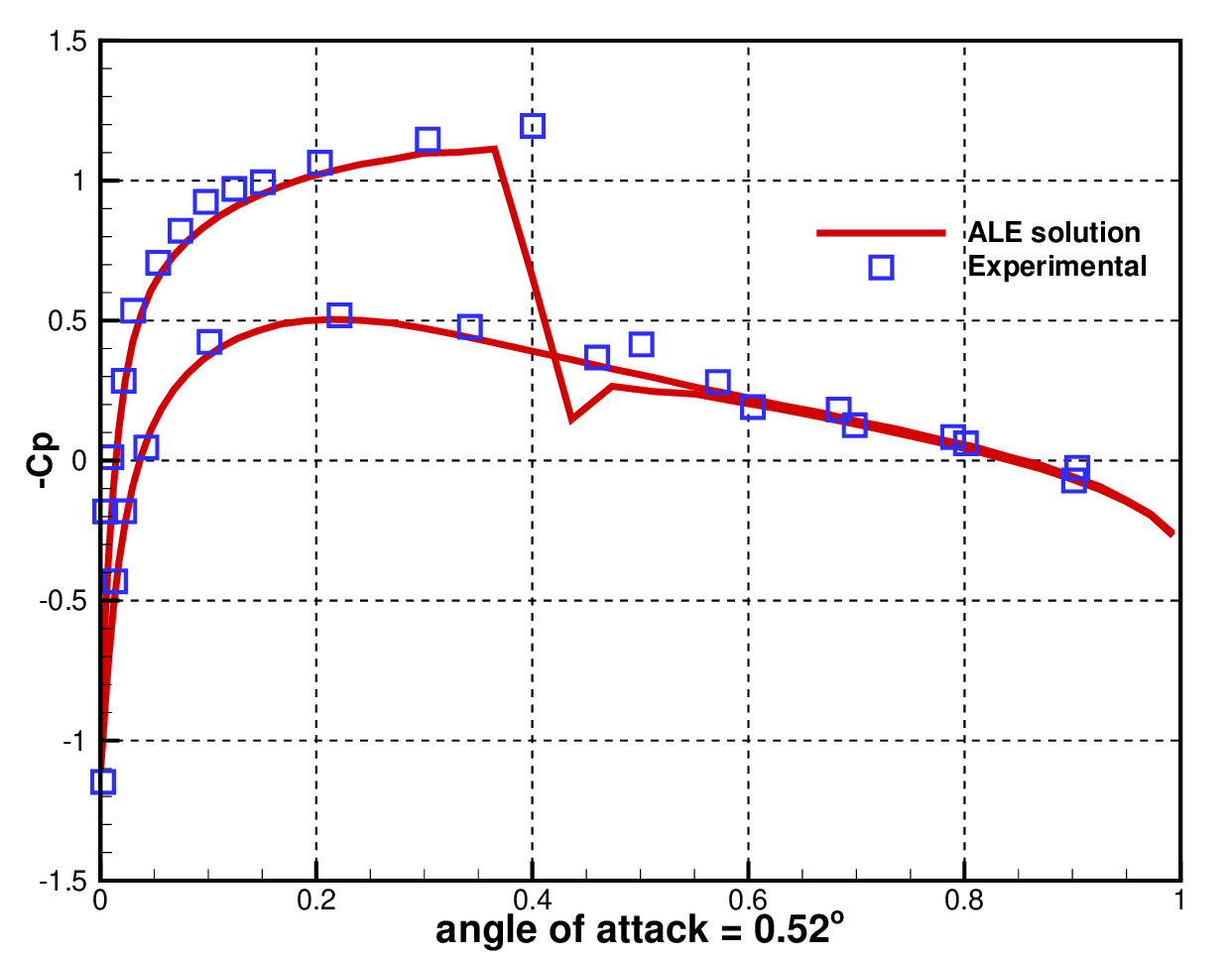} \\
\includegraphics[width=0.45\textwidth]{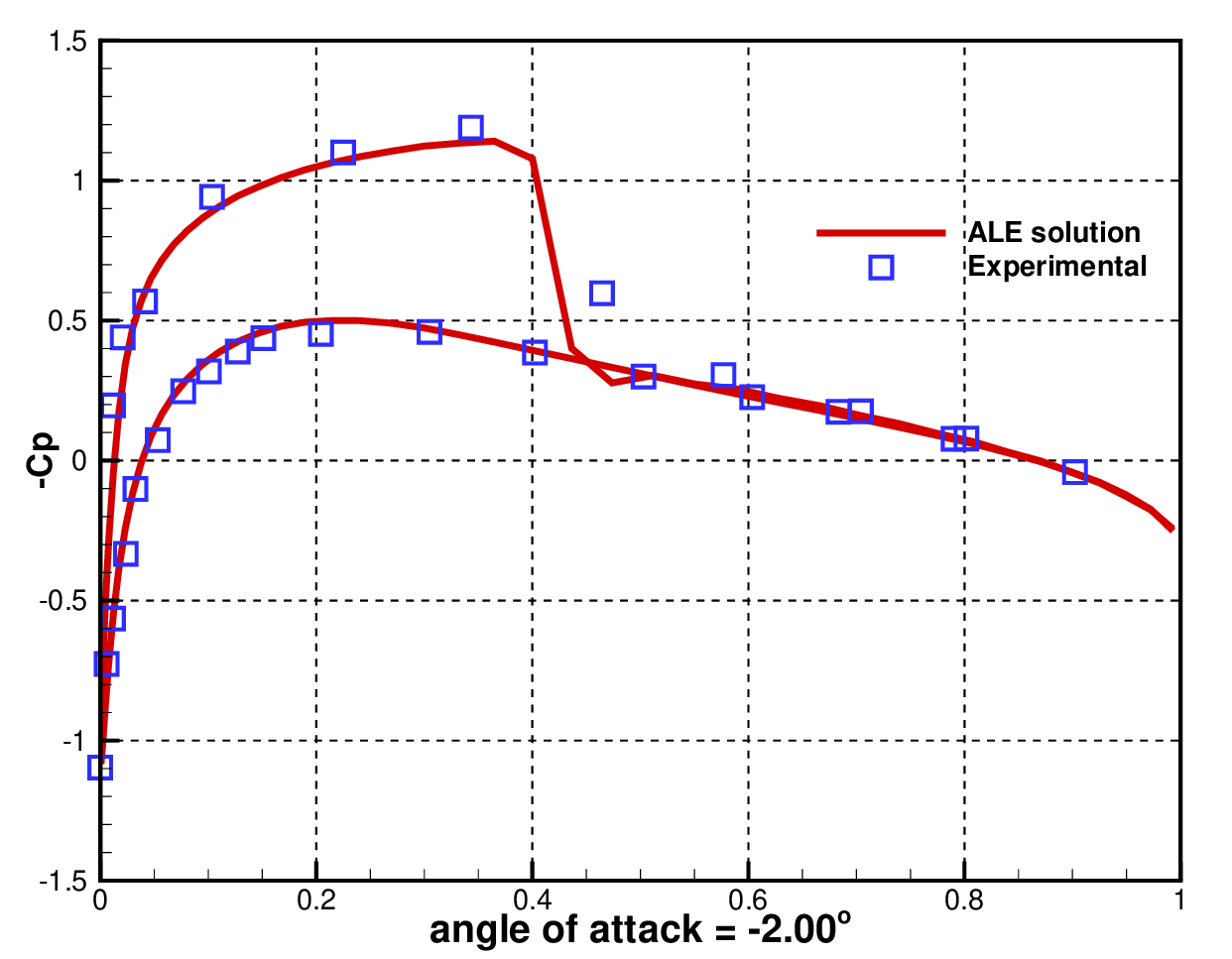}
\includegraphics[width=0.45\textwidth]{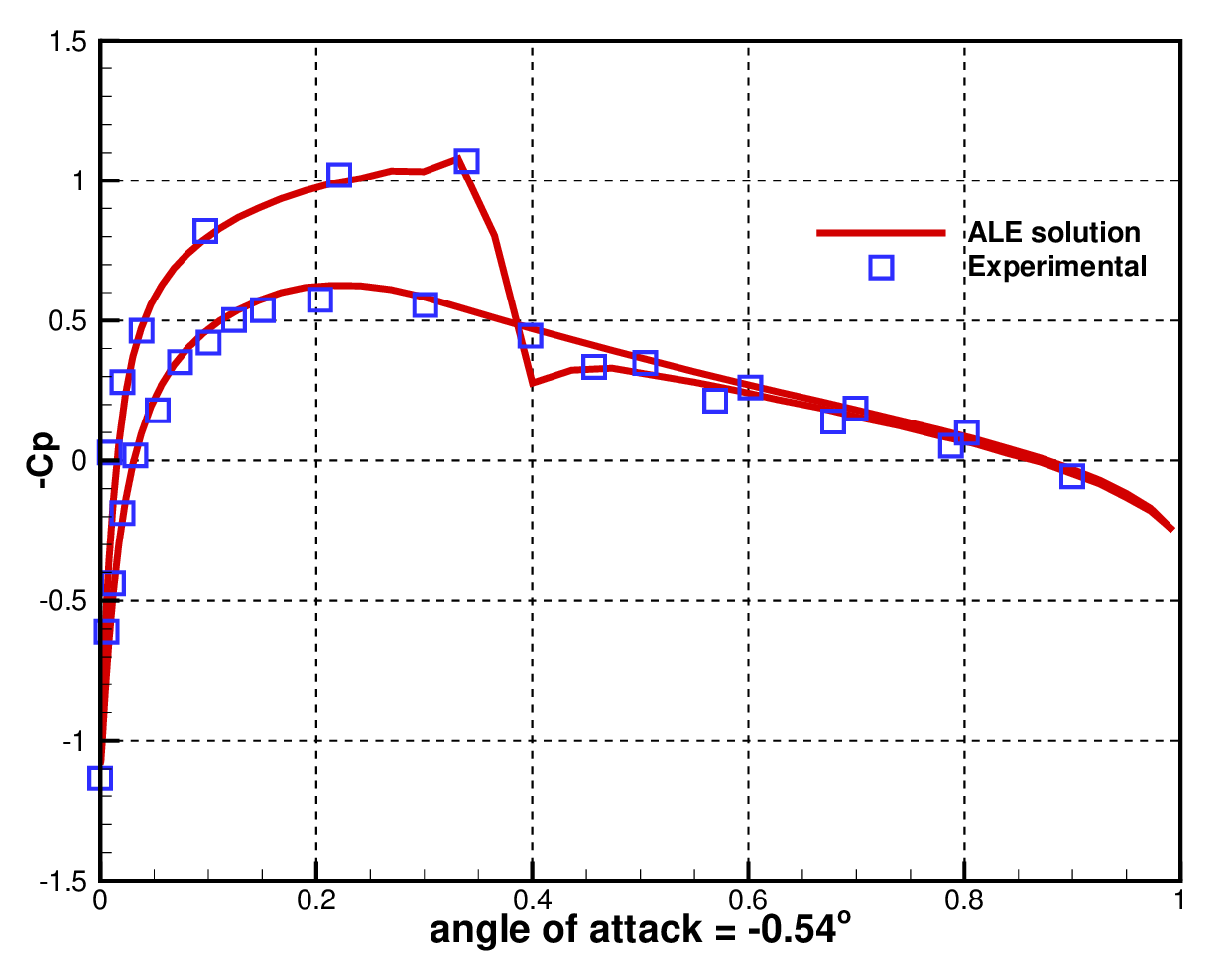}
\caption{\label{airfoli-pre}  Flow over NACA 0012 airfoil:  the pressure distributions on the surface of NACA 0012 airfoil at different angles of attack. }
\end{figure}

In this case, the inviscid transonic flow over NACA 0012 airfoil is tested \cite{Case-airfoil-1, Case-airfoil-2}
The initial angle of attack is $\alpha_m=0.016^\circ$. 
The airfoil is oscillating around a point on the chord, 
which is located at the chord with a quarter of chord length $c$ from the leading point. 
The cycle of oscillation is defined as
\begin{align*}
\alpha=\alpha_m+\alpha_0\sin\omega t,
\end{align*}
where $\alpha_0=2.51^\circ$. The frequency  $\omega$ is given by $\omega c/2U_\infty=0.0814$, 
and the free stream velocity $U_\infty$ is parallel to the $x$-axis and given with the Mach number $0.755$. 
The inviscid boundary condition is used on the surface of the airfoil, 
and the inlet and outlet boundaries are adopted at the outer boundary. 
A C-type mesh with $179\times45$ grid points is used,  
and the mesh around the airfoil oscillates with the given angular velocity in the computation. 
The mesh and pressure distributions at the angle of attack $2.01^\circ$ are presented in Fig.\ref{airfoli-field}. 
The pressure distributions at different angles of attack are given in Fig.\ref{airfoli-pre} with the experiment measurements. 
The numerical results agree well with the experiment data \cite{Case-airfoil-1}. 

\section{Conclusion}
In this paper,  an efficient and robust high-order compact ALE gas-kinetic scheme is developed for the compressible moving grids and moving boundary problems. 
The new memory reduction method is adopted for compact reconstruction on both structured and unstructured meshes. 
In the reconstruction process, a quadratic polynomial can be constructed at the target cell with simply three least square process. 
The two-stage fourth-order method is used for the temporal discretization, 
and the second-order gas-kinetic solver is used for the flux calculation. 
To suppress the spurious numerical oscillations, the gradient compression factor is used as limiter at the cell interface. 
With the new reconstruction and limiting process, the current scheme avoids the large number of matrix inversion operations in traditional high-order ALE schemes, which reduces the memory cost and accelerates  the computation much. 
Numerical examples from smooth flows to the flows with strong discontinuities are presented 
to validate the accuracy, efficiency, robustness, and preservation of geometric conservation law of current scheme. 
In the numerical tests, the mesh velocity is provided by mesh adaptation method and cell centered Lagrangian method. 
Moreover, compared with previous ALE high-order GKS method, a 7x speedup can be achieved, which validates the efficiency of current scheme as well. 
In the future, the method will be extended for three-dimensional problems with implicit schemes to further accelerate the computation. 
Meanwhile, robust and efficient limiters based on GKS formulation will be constructed.

\section*{Acknowledgements}
The current research of L. Pan is supported by Beijing Natural Science Foundation (1232012) and National Natural Science Foundation of China (12494543, 12471364).  
The current research of X. Ji is supported by Funding of National Key Laboratory of Computational Physics, National Natural Science Foundation of China (12302378), 
and the GHfund A (ghfund202407012738).

\section*{Data availability}
The data that support the findings of this study are available from
the corresponding author upon reasonable request.

\end{document}